\documentclass[journal]{IEEEtran}
\usepackage{cite}
\usepackage{color,soul}
\usepackage{gensymb}
\usepackage{dsfont}
\ifCLASSINFOpdf
  \usepackage[pdftex]{graphicx}
  \graphicspath{{../pdf/}{../jpeg/}}
  \DeclareGraphicsExtensions{.pdf,.jpeg,.png}
\else
  \usepackage[dvips]{graphicx}
  \graphicspath{{../eps/}}
  \DeclareGraphicsExtensions{.eps}
\fi
\usepackage[cmex10]{amsmath}
\usepackage{amssymb}

\usepackage{euscript}

\usepackage{multirow}
\usepackage{algorithm}
\usepackage[noend]{algpseudocode}
\usepackage{diagbox}
\usepackage{physics}
\usepackage{tikz}

\ifCLASSOPTIONcompsoc
  \usepackage[caption=false,font=normalsize,labelfont=sf,textfont=sf]{subfig}
\else
  \usepackage[caption=false,font=footnotesize]{subfig}
\fi
\usepackage{url}
\begin{document}
\title{A Data-Driven Customer Segmentation Strategy Based on Contribution to System Peak Demand}

\author{Yuxuan~Yuan,~\IEEEmembership{Student Member,~IEEE,}
	Kaveh~Dehghanpour,~\IEEEmembership{Member,~IEEE,}
	Fankun~Bu,~\IEEEmembership{Student Member,~IEEE,}
	and Zhaoyu~Wang,~\IEEEmembership{Member,~IEEE}
\thanks{This work was supported in part by the Advanced Grid Modeling Program at the U.S. Department of Energy Office of Electricity under Grant DE-OE0000875, and in part by the National Science Foundation under Grant ECCS 1929975. (\textit{Corresponding author: Zhaoyu Wang)}

 Y. Yuan, K. Dehghanpour, F. Bu, and Z. Wang are with the Department of
Electrical and Computer Engineering, Iowa State University, Ames,
IA 50011 USA (e-mail: yuanyx@iastate.edu; wzy@iastate.edu).
 }
}
%\markboth{This paper is a preprint submitted to IET GTD. If accepted, the
%copy of record will be available at the IET Digital Library}%
%{Shell \MakeLowercase{\textit{et al.}}: Bare Demo of IEEEtran.cls for Journals}

\markboth{Submitted to IEEE for possible publication. Copyright may be transferred without notice}%
{Shell \MakeLowercase{\textit{et al.}}: Bare Demo of IEEEtran.cls for Journals}



\maketitle

\begin{abstract}
Advanced metering infrastructure (AMI) enables utilities to obtain granular energy consumption data, which offers a unique opportunity to design customer segmentation strategies based on their impact on various operational metrics in distribution grids. However, performing utility-scale segmentation for \textit{unobservable} customers with only monthly billing information, remains a challenging problem. To address this challenge, we propose a new metric, the coincident monthly peak contribution (CMPC), that quantifies the contribution of individual customers to system peak demand. Furthermore, a novel multi-state machine learning-based segmentation method is developed that estimates CMPC for customers without smart meters (SMs): first, a clustering technique is used to build a databank containing typical daily load patterns in different seasons using the SM data of observable customers. Next, to associate unobservable customers with the discovered typical load profiles, a classification approach is leveraged to compute the likelihood of daily consumption patterns for different unobservable households. In the third stage, a weighted clusterwise regression (WCR) model is utilized to estimate the CMPC of unobservable customers using their monthly billing data and the outcomes of the classification module. The proposed segmentation methodology has been tested and verified using real utility data. 
\end{abstract}

\begin{IEEEkeywords}
Customer segmentation, peak load contribution, observability, machine learning
\end{IEEEkeywords}

% \fi
\IEEEpeerreviewmaketitle

\section{Introduction}

Advent of Advanced metering infrastructure (AMI) has facilitated a deeper understanding of customer behaviors in low-voltage networks for distribution system operators. Individual customers' demand consumption can be recorded by smart meters (SMs) with high temporal resolution, which enables developing novel data-centric grid operation mechanisms. One of these mechanisms is utility-scale customer segmentation \cite{RG2015}, which is extremely useful in enhancing system operation and management by intelligently targeting customers for peak shaving programs, AMI investment, and retail price/incentive design. This will help utilities under strict financial constraints to optimize their investment portfolio. However, for small-to-medium utilities, a key barrier against investigating an efficient customer segmentation is the absence of real-time measurements due to financial limitations \cite{yuan2019}. Currently, more than half of all U.S. electricity customer accounts do not have SMs to record their detailed consumption behavior \cite{eiasmart}.

%the widespread deployment of smart meters (SMs) provides a good opportunity to improve the understanding of customer behaviors for distribution system operators.
%The development of smart grid technologies has facilitated a deeper understanding of customer behaviors in low-voltage networks for distribution system operators. According to statistical data, the number of smart meters (SMs) in the U.S. exceeded 70 million by the end of 2016 \cite{eiasmart}. The widespread deployment of SMs provides a good opportunity to reduce the system peak demand by monitoring the real-time customer behaviors. Power system peak demand is typically met by peak generators leading to an overall increase in operation costs due to the higher marginal cost of these generators \cite{peakdemand2011}. In the past, due to the limited number of real-time measurement devices, the customers with high demand levels, such as industrial and large commercial customers, have been the primary targets for utilities to participate in various power programs \cite{MH2008}. More recently, motivated by our utilities partners, a critical task for the majority of small-to-medium utilities is identifying candidate residential customers, which can then be targeted for installing SMs, designing the retailed-rates, and providing additional flexibility for peak reduction. This will help utilities under strict financial constraints to optimize their benefits.

Several papers have focused on developing customer segmentation strategies using SM data. One of the most common approaches is to leverage clustering techniques for identifying typical load profiles \cite{cluster2013, cluster2016,RL20152}. In \cite{cluster2013}, principal component analysis (PCA) is performed to extract the dominant features within customer consumption data and then k-means algorithm is employed to classify consumers. In \cite{cluster2016}, a finite mixture model-based clustering is presented to obtain distinct behavioral groups. In \cite{RL20152}, a C-vine copulas-based clustering framework is proposed to carry out consumer categorization. However, the typical load profile extraction alone is insufficient to assess customers' impacts on system peak demand, which limits utilities' ability to target suitable customers for reducing the operation costs.

Apart from typical load profiles, several customer segmentation methodologies have been developed based on the feature characterization and extraction \cite{AK2015, JK2014, good2018, MYS2019}. In \cite{AK2015}, residential customers are ranked using their appliance energy efficiency to reduce building energy consumption. In \cite{JK2014}, the entropy of household power demand is used to evaluate the variability of consumption behavior, which is considered to be a key component in peak shaving program targeting and customer engagement. In \cite{good2018}, a customer's marginal contribution to system cost is obtained using daily demand profiles. In \cite{MYS2019}, a four-stage data-driven probabilistic method is proposed to estimate the coincident peak demand estimation of new customers for designing new systems. Compared to the clustering approaches, these methods directly quantify customer-level features from SM data and use them to determine the segmentation strategies. Nevertheless, the previously-proposed metrics fall short of considering customers' impact on system peak demand, which is a major problem considering that continuous growth in system peak load raises the possibility of power failure and increases the marginal cost of supply \cite{pss}. Furthermore, previous works have only focused on observable customers. 

%Thus, it remains a big challenge to find the high-impact unobservable customers for heterogeneous power programs by estimating the peak contribution of individual load. However, more than half of all U.S. electricity customer accounts do not have SMs to record their detailed consumption behavior currently \cite{eiasmart}. For example, it cannot be claimed that a low entropy household has a higher contribution to the system peak demand compared to a customer with a more variable consumption profile. M

In order to address these shortcomings, this paper proposes a new metric for customer segmentation, which is denoted as coincident monthly peak contribution (CMPC). CMPC is defined as the ratio of individual customer's demand during system daily peak load time over the real-time total system peak demand in a course of a month. Compared with conventional coincident peak demand metrics, which quantify the peak consumption levels of multiple customers based on their empirical diversified maximum demand \cite{MYS2019}, the proposed CMPC focuses on the impact of individual customer and conveys information on how individual customer's peak time differs from the system's peak demand time. Based on the definition of CMPC, we develop a multi-stage machine learning-based customer segmentation strategy that estimates CMPCs of unobservable customers using only their monthly billing information. The developed method consists of three modules: 1) Using a graph theoretic clustering, a seasonal typical load pattern bank is constructed to classify various customer consumption behaviors. 2) To connect unobservable customers to the seasonal databank, a multinomial classification model is presented which identifies typical load profiles of customers without SMs. 3) According to the outcome of the classification module, a weighted clusterwise regression (WCR) model is trained to map the unobservable customers' monthly energy consumption data to CMPC values. 
%In \cite{km2019}, \hl{a conceptually-similar three-stage framework is developed to perform imbalance-induced energy losses for data-scarce distribution systems using the cumulative density functions (CDFs). Compared to this work, we focus on a distinct goal of quantifying the peak contribution for the unobservable customers at the grid-edge. The main challenge is the higher uncertainty of customer-level load, which increases the difficulty for pattern construction and CMPC inference.}
%This machine learning framework takes advantage of the considerable correlation between CMPCs and monthly consumption levels for different customer profiles. 
Utilizing our segmentation method, within a certain range of consumption, customers with heavy demand but small contribution to the system peak could be excluded from AMI investment/peak shaving investment portfolios, whereas those with a similar demand level but a larger peak contribution can be targeted in such programs as impactful customers. The main contributions of this paper can be summarized as follows: 
\begin{itemize}
\item A new metric, CMPC, is proposed as a measure for customer segmentation strategy, which accurately assesses the individual customer impact on system peak from a real dataset. We will show that the proposed metric contains different and unique information compared to the existing metrics. 
\item A three-stage machine learning framework is developed to obtain CMPC for unobservable customers by accurately estimating their contribution to system peak demand.
\item The proposed framework is innovative and intuitive, and considers various specific properties of our real data: 1) the linear nature of the relationship between the CMPC and demand level in the same cluster; 2) concentration of residential customers demand within a small range; 3) strong seasonal changes in customer behaviors.
\item The proposed framework can handle the uncertainty of the classification process by integrating the probabilistic values for each typical pattern in the regression model. 
\end{itemize}

%The rest of the paper is organized as follows: Section \ref{overall} describes the real data and the proposed CMPC metric. In Section \ref{cluster}, a clustering algorithm is presented to build the seasonal consumption pattern bank. In Section \ref{WCR}, CMPC inference for unobservable customers is proposed using a classification model and WCR. The numerical results are analyzed in Section \ref{result}. Section \ref{conclusion} presents research conclusions.

\section{Data Description and CMPC Definition\label{overall}}
\subsection{Data Description}

\begin{figure}[tbp]
\centering
\includegraphics[width=3.5in]{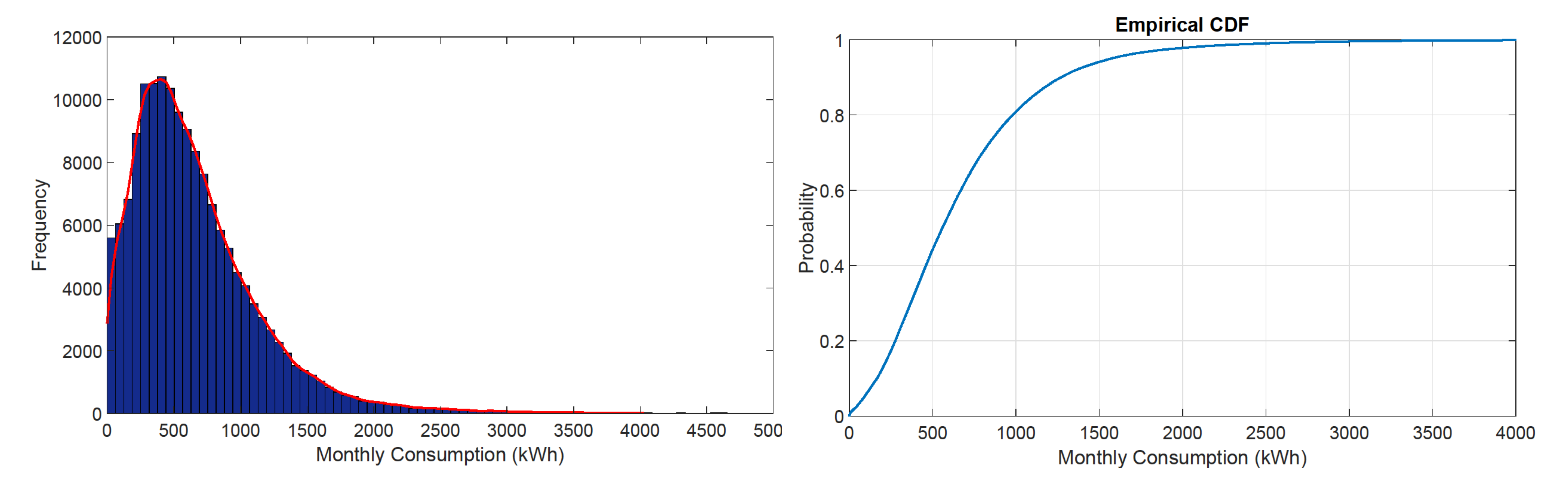}
\caption{Monthly consumption distribution: consumption histogram (left), consumption CDF (right).}
\label{fig:energy_dis}
\end{figure}

The available data used in this paper is provided by several mid-west U.S. utilities. The data includes the energy consumption measurements of over 3000 residential customers from SMs, and the corresponding supervisory control and data acquisition (SCADA) data. The data ranges from January 2015 to May 2018 \cite{wzy}. The SM data was initially processed to eliminate grossly erroneous and missing samples. Accordingly, the data points with a z-score magnitude of larger than 5 are marked as ``erroneous'' and replaced using local interpolation \cite{zscore}. The empirical distribution and cumulative distribution function (CDF) of customer monthly energy consumption are obtained and presented in Fig. \ref{fig:energy_dis}. As shown in the figure, the majority of residential customer monthly consumption samples are concentrated around 1000 kWh, and almost 80$\%$ of customers have monthly consumption levels below 1000 kWh. Compared to the industrial and commercial customers, the demand level of residential households is distributed within a smaller range. This indicates that using only demand level for customer segmentation can be a difficult task.
\begin{figure}[tbp]
\centering
\includegraphics[width=3.5in]{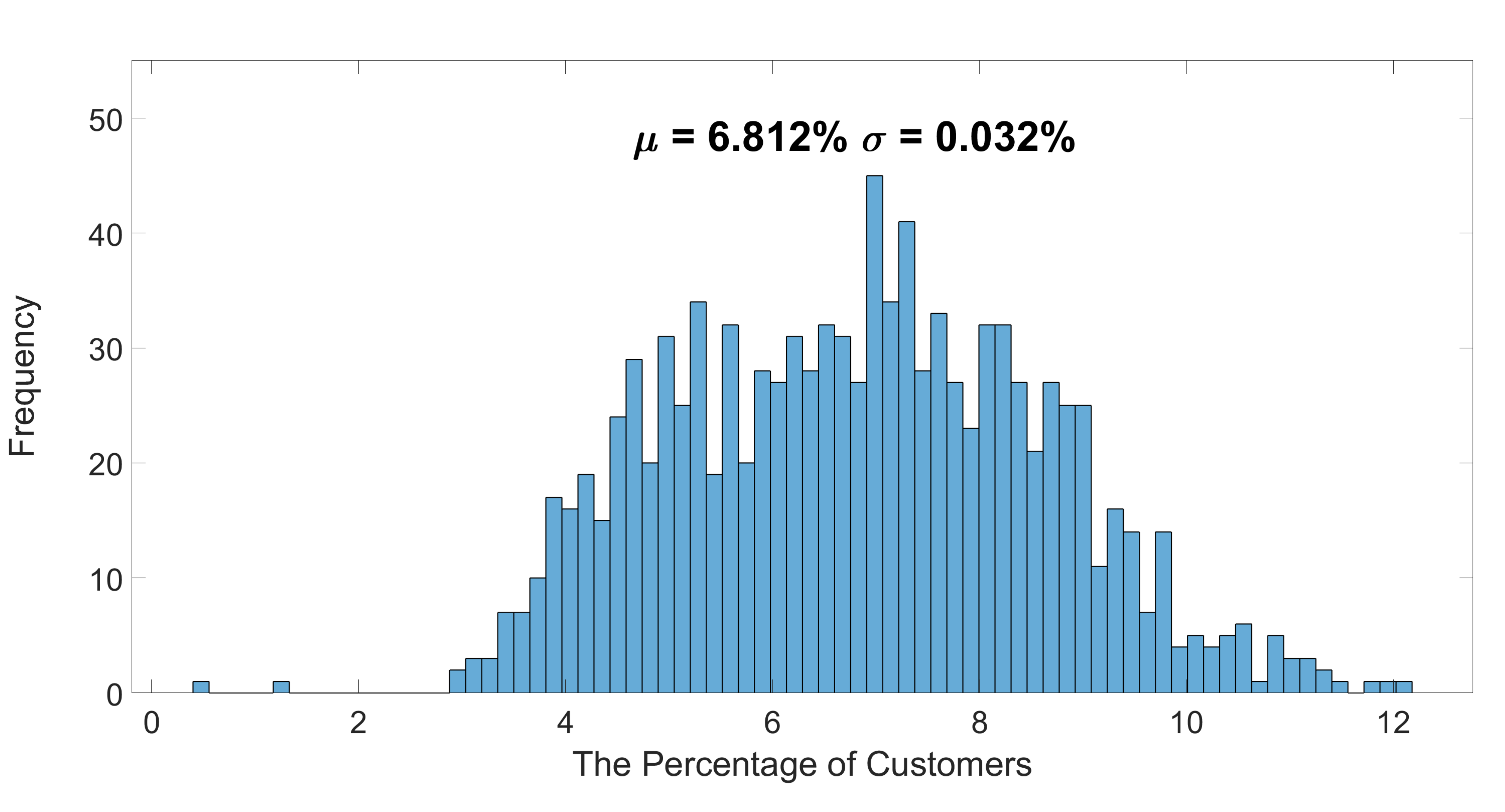}
\caption{Percentage of customers whose peak demand coincide with the system peak.}
\label{fig:customer_peak_system_peak}
\end{figure}

\subsection{CMPC Definition}

%\begin{figure}[htbp]
%\centering
%\includegraphics[width=3.5in]{figure/sample_seasons_average_load_different.pdf}
%\caption{Seasonal average load profiles of a sample customer.}
%\label{fig:sample}
%\end{figure}
The system peak demand is one of the most important operational factors for utilities due to the high marginal cost of energy procurement at the peak time. Hence, it is obligatory to investigate a customer segmentation methodology based on each load's contribution to system peak demand. However, individual customer's peak demand cannot be employed as a measure to assess this contribution, since individual customer peak demand does not necessarily coincide with the system peak. In order to illustrate this, a statistical analysis is performed on the available SM dataset. Fig. \ref{fig:customer_peak_system_peak} shows the percentage of customers whose peak demand coincides with the system peak load. On average only 6$\%$ of customers have the same peak time as the system, with a standard deviation of 12$\%$. This means that a customer's peak demand cannot be relied upon to estimate its contribution to the overall system peak load. Thus, in this paper, we propose a new metric, denoted as CMPC, to accurately quantify the contribution of an individual customer to the system peak demand:
\begin{figure}[htbp]
\centering
\includegraphics[width=3.5in]{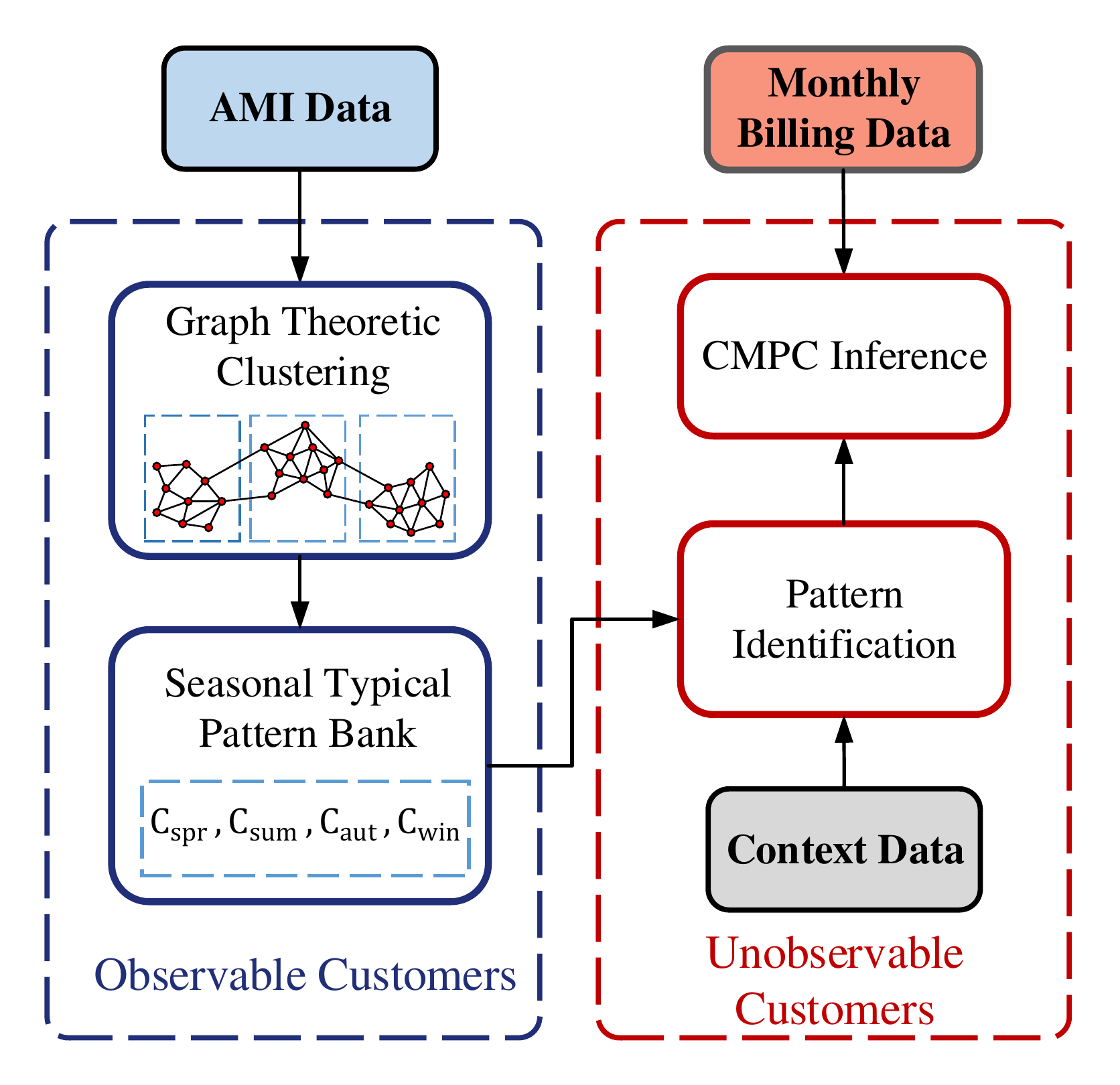}
\caption{Proposed data-driven framework.}
\label{fig:main}
\end{figure}
\begin{equation}
\label{eq:CMPC}
F_{j,m} = \frac{1}{n}\sum_{d=1}^n\frac{p_{j,m}^d(t_{d})}{P_m^d(t_{d})}
\end{equation}
where CMPC of the $j$'th customer at the $m$'th month is denoted by $F_{j,m}$. Here, $p_{j,m}^d(t_{d})$ is the customer's demand at time $t_{d}$ on the $d$'th day of the month, with $n$ denoting the total number of days in the month. Note that $P_m^d$ and $t_{d}$ are the value and the time of system peak demand on the $d$-th day of the $m$-th month. Hence, CMPC is basically the average customer contribution to the daily system peak demand during a month. A few related but different indices can be found in the literature, such as \textit{coincidence contribution factor}, which is defined as the gap between the aggregate peak demand of a group of customers and their actual consumption at the system peak time \cite{RL2015}. However, the coincidence contribution factor cannot be used as a customer-level metric due to its inability to quantify individual customers' contributions to the system peak load. 
\begin{figure}[tbp]
\centering
\subfloat [Summer statistics]{
\includegraphics[width=3in]{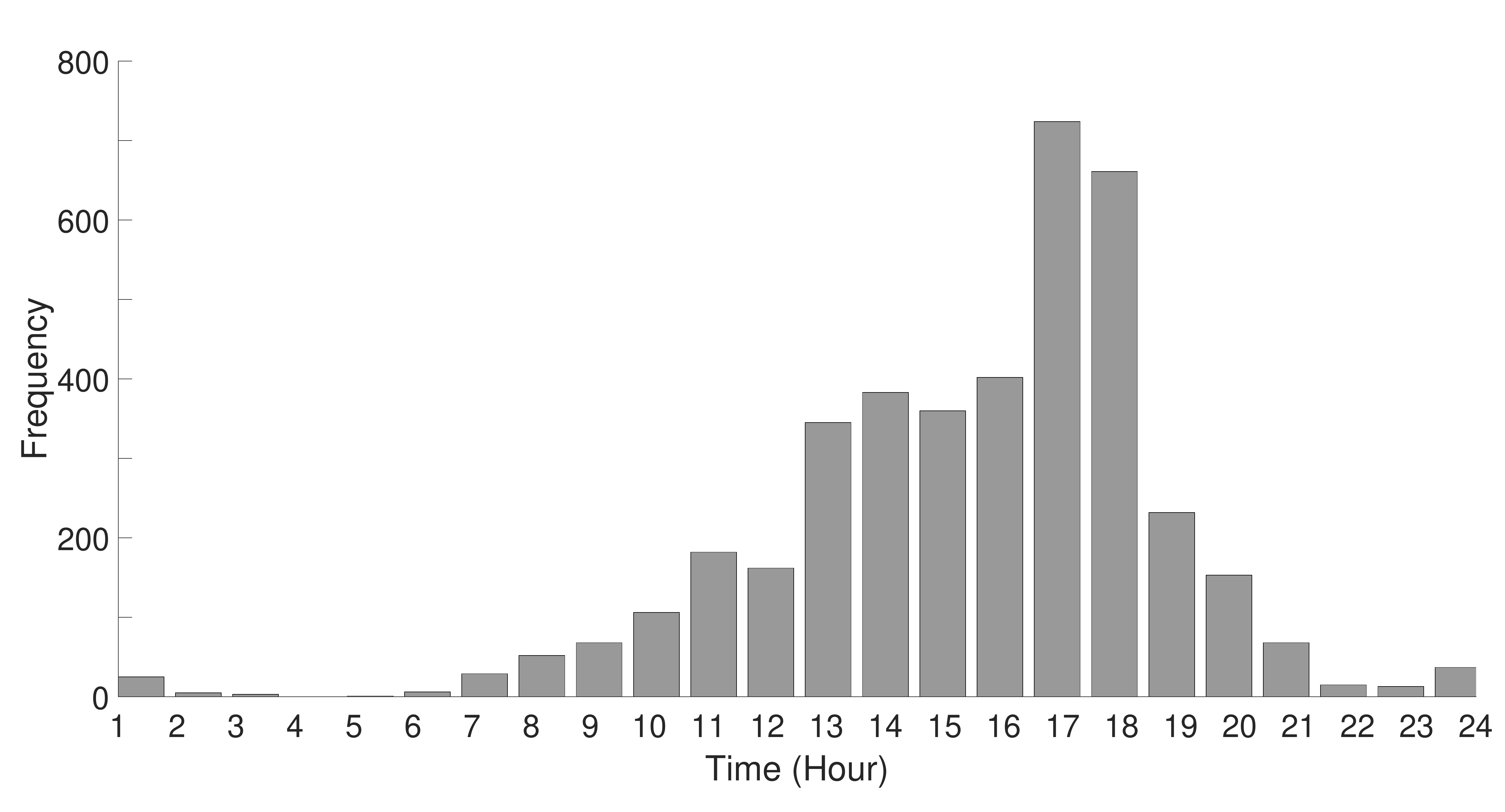}
}
\hfill
\subfloat [Winter statistics]{
\includegraphics[width=3in]{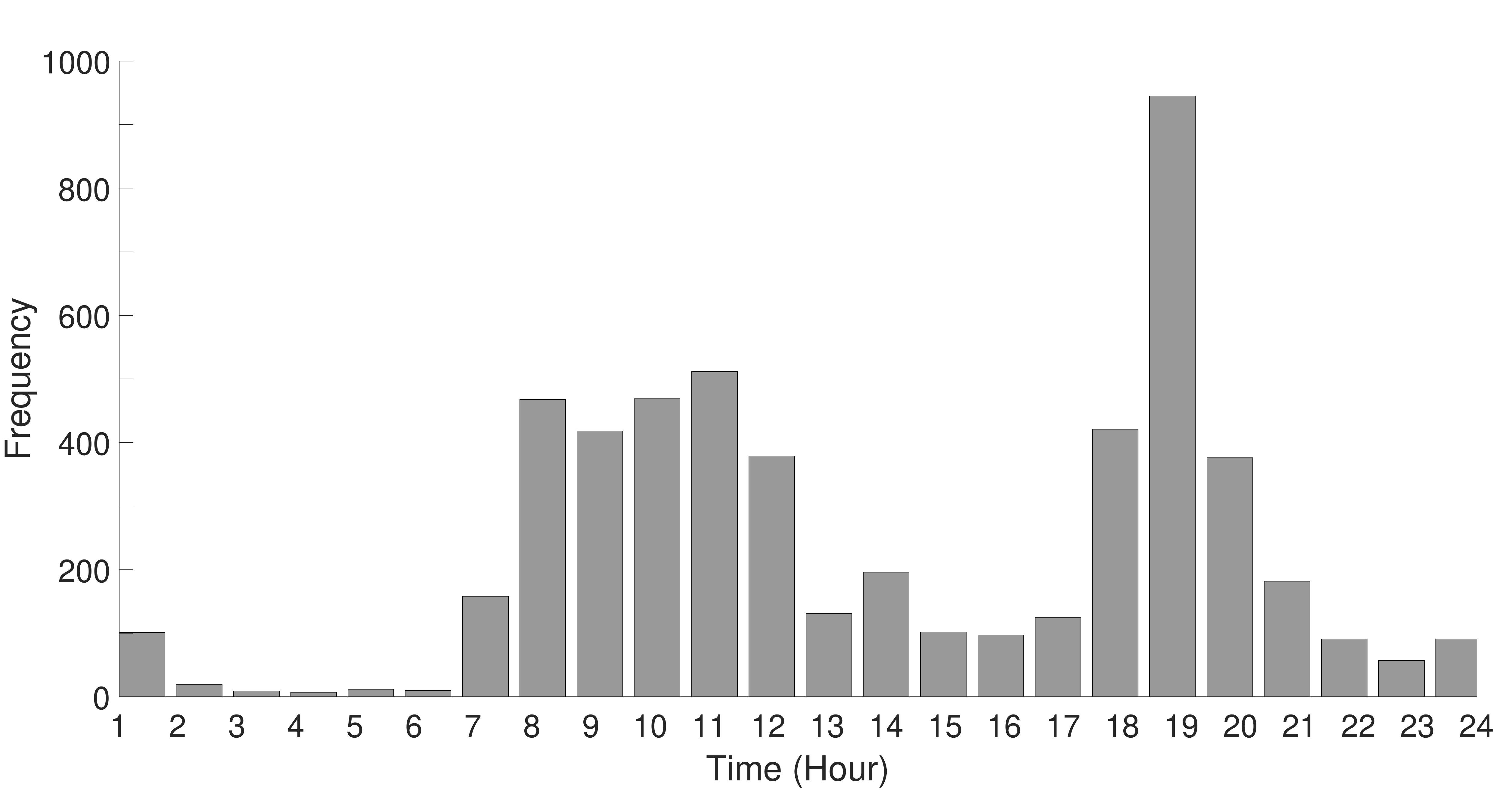}
}
\caption{Seasonal system peak time distribution.}
\label{fig:season}
\end{figure}

CMPC can be directly calculated for observable customers using the real-time SM measurements. Considering that not all customers have SMs in practice, especially for residential households, we propose a multi-stage data-driven method for estimating CMPC. The flowchart of the proposed approach is presented in Fig. \ref{fig:main}. (I) In the first stage, the demand profiles of observable customers are utilized to build a seasonal consumption pattern bank, $[\{C_{spr}\},\{C_{sum}\},\{C_{aut}\},\{C_{win}\}]$, using a graph theoretic clustering technique. Here, each $\{C_{(\cdot)}\}$ is the set of the typical daily load profiles for a specific season (detailed in Section \ref{cluster}). Seasonal data clustering shows a better load behavior identification performance due to its ability to capture the critical seasonal behaviors of customers \cite{KC2017}. (II) Then, a classification module is developed to infer the likelihood of identified seasonal daily consumption profiles for customers without SM data utilizing sociodemographic information. (III) For each typical pattern, a regression model is trained to provide an inference function to estimate the CMPC from customers' monthly billing data. To take into account the variances of CMPC in different typical patterns, a WCR approach is developed based on the results of classification module. Basically, the proposed customer segmentation approach is able to infer CMPC of customers without SMs using their monthly billing information and limited context information.

\section{Graph Theoretical Clustering Algorithm}\label{cluster}

In this paper, a graph theory-based clustering technique, known as spectral clustering (SC), is adopted. Due to the strong seasonal changes in the customers' behavior, the SC uses seasonal average customer load profiles to identify typical daily load patterns corresponding to different seasons \cite{NMW2002, kaveh2019}. According to the statistical analysis, both customer behaviors and system peak timing are affected by seasonal changes, as shown in Fig. \ref{fig:season}. In Fig. \ref{fig:season}$(a)$, the peak time distribution in summer is concentrated around evening interval (17:00-18:00 pm). Meanwhile, the peak time probability rises during daytime and falls sharply at night. One possible reason is the increase of air conditioning usage during summer daytime. In contrast, the peak time distribution of winter is presented in Fig. \ref{fig:season}$(b)$. Compared to the summer, the distribution of peak demand time in winter has two concentration points: one in morning hours (8:00-12:00 am), and the other in the evening (18:00-20:00 pm). Also, the peak time probability shows relatively low values during the afternoon interval (13:00-17:00 pm). Hence, in this work, instead of assigning a single pattern to each customer, various patterns are obtained for different seasons to capture the seasonality of customer behaviors \cite{KC2017}. % The details of the SC algorithm are shown in Algorithm \ref{alg:SC}. 

In each season, the AMI dataset is represented as an undirected similarity graph, $G=(V,E)$. $V$ is the set of vertices in the graph, where the $i$'th vertex represents the average daily profile of the $i$'th customer, $V_i=[C_{1}^i,...,C_{24}^i]$, with $C_{j}^i$ denoting the average load value at the $j$' hour of day for the $i$'th customer. $E$ is the set of edges in the graph that connect different vertices, where a non-negative weight, $W_{i,j}$, is assigned to the edge connecting vertices $i$ and $j$. The weight value represents the level of similarity between the two customers' average daily load profiles, with $W_{i,j}=0$ indicating that the vertices $V_i$ and $V_j$ are not connected. In this paper, the weight $W_{i,j}$ is obtained by adopting a Gaussian kernel function:
\begin{equation}
\label{eq:adjacency}
W_{i,j} = exp(\frac{-||V_i-V_j||^2}{\alpha^2})
\end{equation}
where $\alpha$ is a scaling parameter that controls how rapidly the weight $W_{i,j}$ falls off with the distance between vertices $V_i$ and $V_j$. To enhance computational efficiency and adaptability to the dataset, we have adopted a localized scaling parameter $\alpha_i$ for each vertex that allows self-tuning of the point-to-point distances based on the local distance of the neighbor of $V_i$ \cite{GJ2007}: 
\begin{equation}
\label{eq:local_scale}
\alpha_i = ||V_i-V_\varphi||
\end{equation}
where, $V_\varphi$ is the $\varphi$'th neighbor of $V_i$, which is selected according to \cite{GJ2007}. Therefore, the weight between a pair of points can be re-written as:
\begin{equation}
\label{eq:adjacency_1}
W_{i,j} = exp(\frac{-||V_i-V_j||^2}{\alpha_i\alpha_j})
\end{equation}
Given a set of vertices and weight matrix $W=(W_{i,j})_{i,j=1,...,n}$, the clustering process is converted to a graph partitioning problem. In this paper, the objective function of graph partitioning is to maximize both the dissimilarity between the different clusters and the total similarity within each cluster \cite{ISD2007}:
\begin{equation}
\label{eq:ncut1}
N(G) = \min_{A_1,...,A_n}\sum_{i=1}^{n}\frac{c(A_i,V\setminus {A_i})}{d(A_i)}
\end{equation}
where, $n$ is the number of vertices, $A_i$ is a cluster of vertices in $V$, $V\setminus A_i$ represents the nodes of set $V$ that are not in set $A_i$, $c(A_i,V\setminus A_i)$ is the sum of the edge weights between vertices in $A_i$ and $V\setminus A_i$, $d(A_i)$ is the sum of the weights of vertices in $A_i$. It has been shown in \cite{NMW2002} that the minimum of $N(G)$ is reached at the second smallest eigenvector of the graph's Laplacian matrix, $L$, which can be determined using the weight matrix $W$, as demonstrated in:
\begin{equation}
\label{eq:laplaican}
L = D^{-\frac{1}{2}}WD^{-\frac{1}{2}}
\end{equation}
where, $D$ is a diagonal matrix, which $(i,i)$'th element is the sum of $W$'s $i$'th row. The $k$ smallest eigenvalues, $[y_1,y_2,...,y_k]$, of the Laplacian matrix are extracted in the clustering algorithm (see Alg. 1) to build a new matrix $U \in\mathbb{R}^{n\times{k}}$, where $k$ ranges from $2$ to $n$. Leveraging the properties of the graph Laplacians, the data point $V_i$ is reconstructed using the $i$'th row of the $U$ matrix, which enhances the cluster-properties of the data \cite{GJ2007}. After data reconstruction, a simple clustering algorithm is able to detect the clusters. In this work, we utilized the $k$-means algorithm to obtain the final solutions from matrix $U$.

\begin{figure}[tbp]
\centering
\includegraphics[width=3.2in]{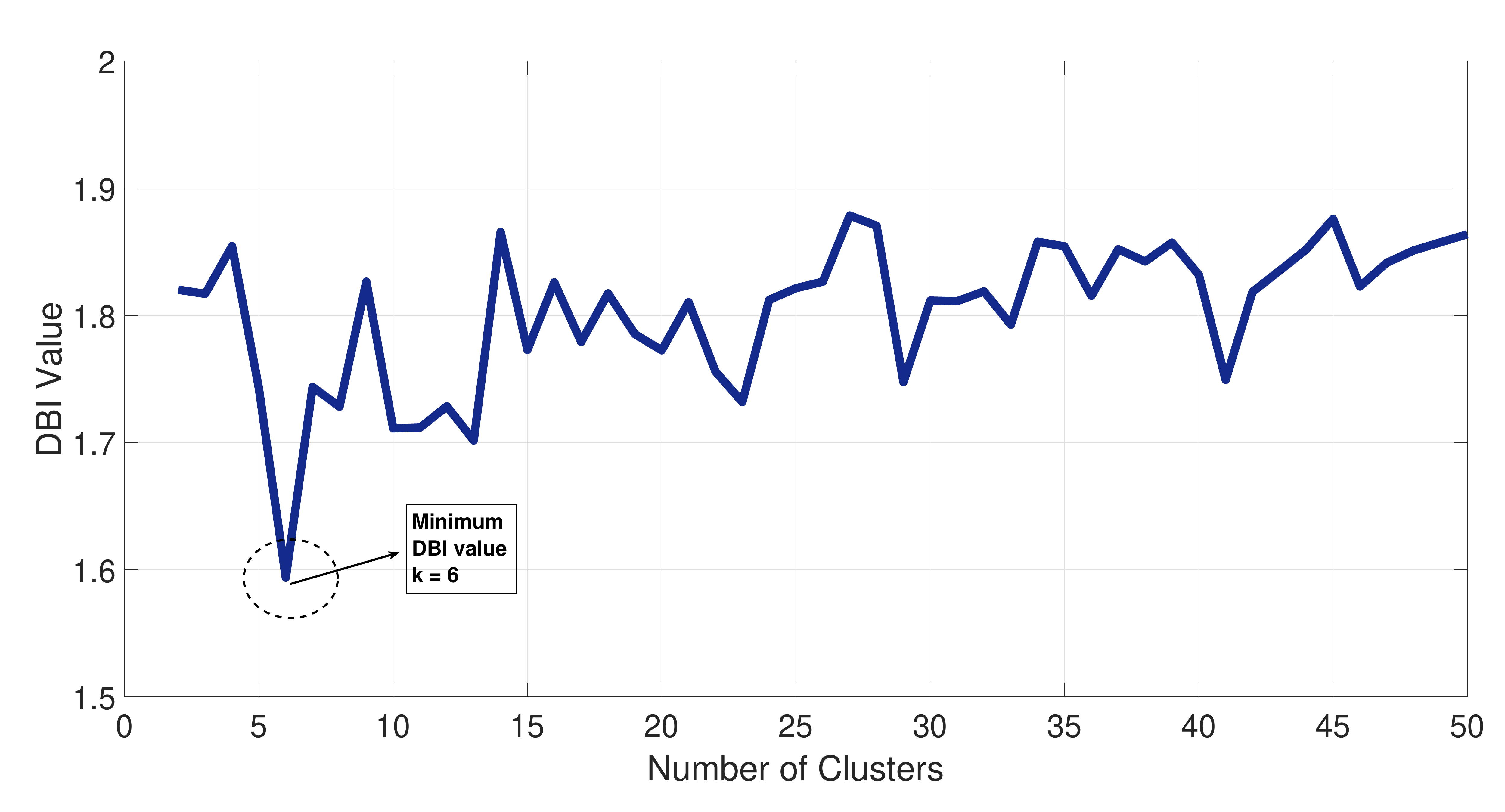}
\caption{Cluster validation index performance for summer season.}
\label{fig:DBI}
\end{figure}

Compared to conventional clustering techniques, the SC algorithm has two main advantages: $(1)$ it mainly relies on the weight matrix of the dataset rather than using the high-dimensional demand profile data directly. Also, computing the eigenvalues of matrix $W$ for data reconstruction is equivalent to achieving dimension reduction by employing a linear PCA in a high dimensional kernel space; $(2)$ as a basic idea of SC, graph partitioning problem can be solved without making any assumptions on the data distribution. This improves the robustness of SC, and leads to better clustering performance for complex and unknown data structures \cite{GJ2007}. $(3)$ According to equations \ref{eq:adjacency}-\ref{eq:laplaican}, SC converts the clustering process to a graph partitioning optimization problem. Based on \textit{Rayleigh-Ritz theorem}, the solution of this optimization problem is obtained using the $k$ eigenvectors of the Laplacian matrix, which guarantees a good approximation to the \textit{optimal cut}. \cite{lap1997, DS1996, CC2013} The main challenge of SC is that the $k$ value still needs to be determined as a priori. To obtain the optimal $k$, we employ the Davies-Bouldin validation index (DBI), which aims to maximize the internal consistency of each cluster and minimize the overlap of different clusters \cite{DV2016}. The optimal value of $k$ can be obtained when the DBI is minimized. This is shown in Fig. \ref{fig:DBI} for summer data subset.

%\IEEEpubidadjcol

\section{CMPC Estimation for Unobservable Customers}\label{WCR}
\begin{figure*}[tbp]
      \centering
      \includegraphics[width=2\columnwidth]{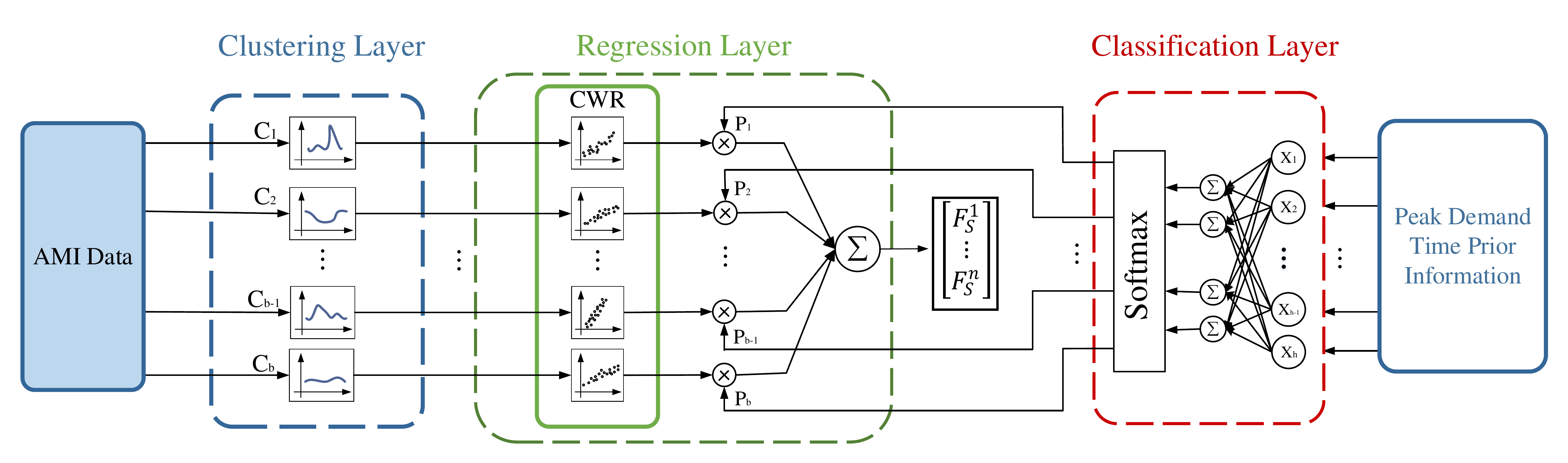}
\caption{The structure of WCR model.}
\label{fig:draft}
\end{figure*}  
In order to assess the CMPC of unobservable customers, a WCR approach is proposed using only their monthly consumption information, as shown in Fig. \ref{fig:draft}. This framework includes two stages: the first stage is unobservable customer classification based on the seasonal typical consumption pattern bank, and the second stage is cluster-based CMPC inference. It should be noted the two stages cannot be directly combined into one step since they address two different problems.
\begin{figure}[tbp]
\centering
\subfloat [Monthly energy and CMPC of different patterns in spring]{
\includegraphics[width=3in]{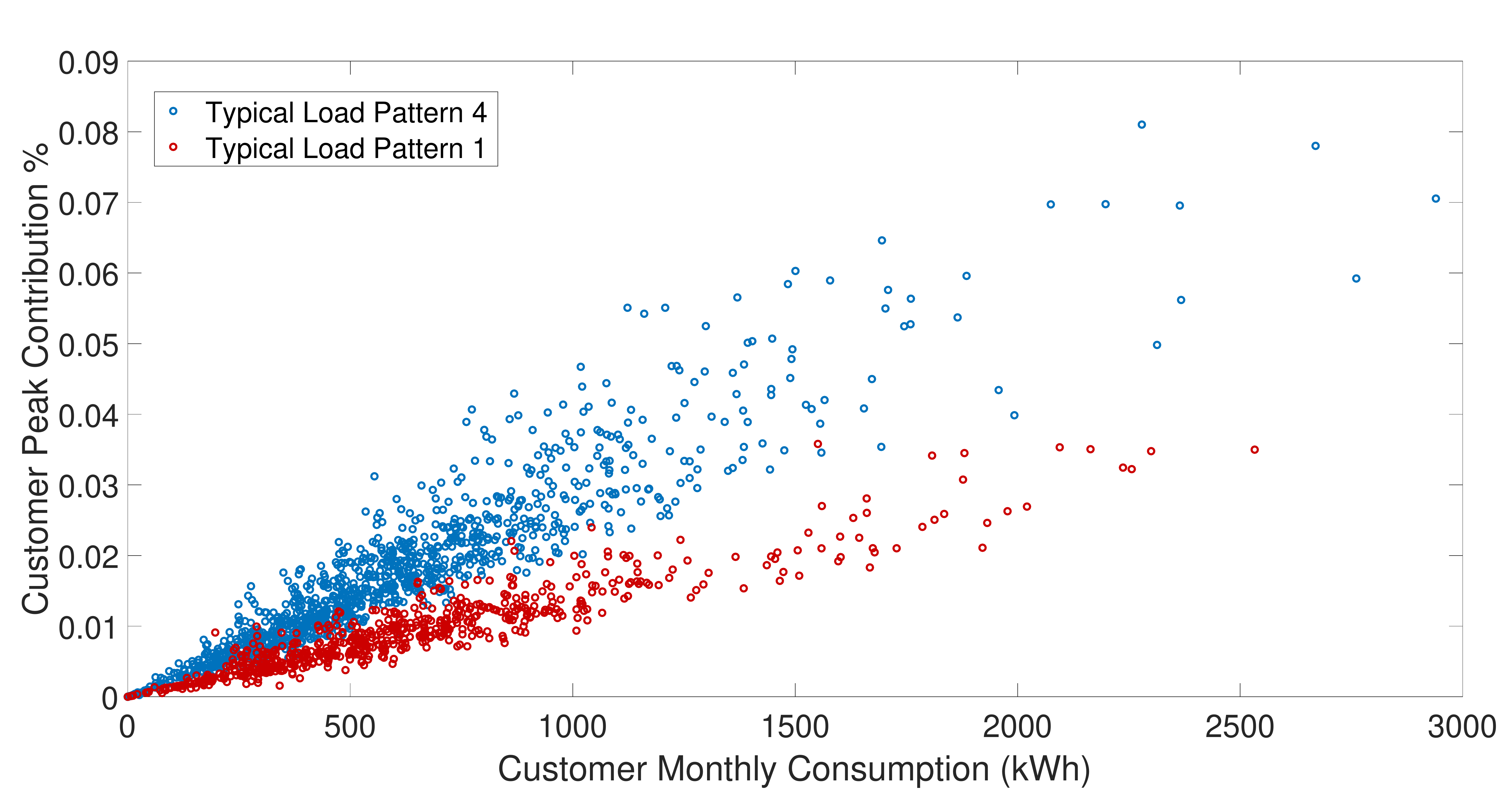}
}
\hfill
\subfloat [Monthly energy and CMPC of different patterns in summer]{
\includegraphics[width=3in]{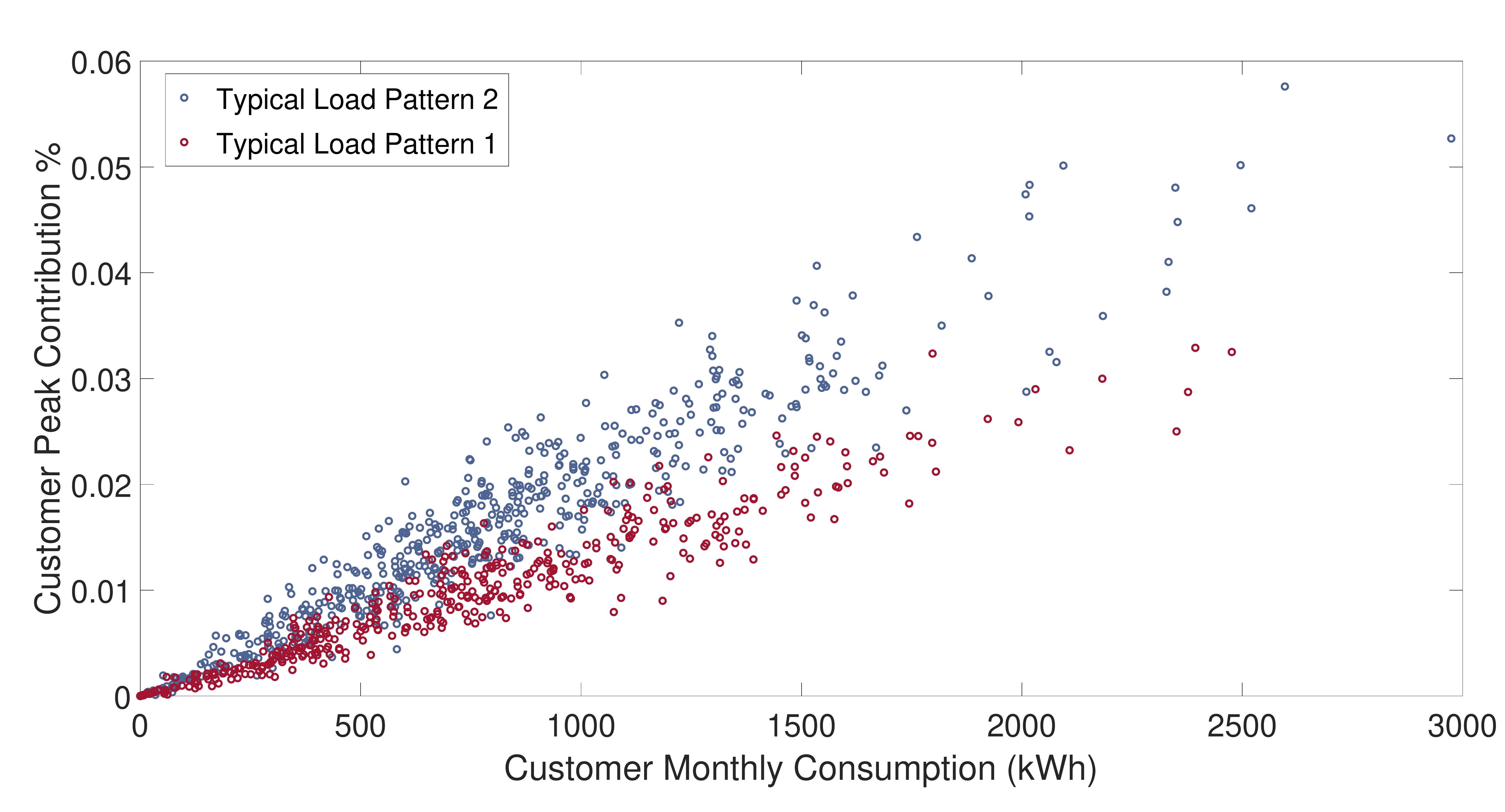}
}
\caption{Performance of clusterwise.}
\label{fig:linear}
\end{figure}

\subsection{Unobservable customer classification}
Since the detailed time-series SM data of unobservable customers is not available, their daily consumption patterns cannot be directly determined beforehand. To link the existing typical load patterns, obtained from the SC technique, to unobservable customers, a pattern classification model is developed. Thus, the goal of this model is to design a classifier that is able to distinguish different behavioral classes based on an input vector that contains sociodemographic information of unobservable customers. The proposed model in this paper maps the sociodemographic information of customers (i.e. working period and dining time) to the typical daily pattern databank. The basic idea is that the typical daily load profiles of customers can be discovered using prior knowledge of their peak consumption timing. 

Based on the sociodemographic information of customers, the knowledge of customer behavior over a few distinctive intervals in the day can be obtained, namely the morning interval (from 7:00 am to 9:00 am), the afternoon interval (from 12:00 pm to 14:00 pm), and the evening interval (from 18:00 pm to 21:00 pm). This prior information is then used to obtain an approximate probability distribution function of customer peak timing defined as $X^j = \{X_1^j,X_2^j,...,X_{h-1}^j,X_h^j\}$, where $X_i^j$ is the probability of $j$'th customer peak demand occurring at time instant $i$, with $h$ denoting the maximum number of time points. In this work, using the SM measurements of observable customers, $X_i^j$ is determined as follows:
\begin{equation}
\label{eq:C1}
X_i^j = \frac{\sum_{d=1}^n \Phi(t_d^{j})}{n}
\end{equation}
\begin{equation}
\label{eq:C2}
\Phi(t_d^{j}) = \begin{cases}
1 &\mathrm{for} \ t_d^j = i\\
0 &\mathrm{for} \ otherwise
\end{cases} 
\end{equation}
where, $t_d^j$ is the peak demand time of $j$'th customer at the $d$-th day. Thus, the peak timing likelihood distribution, $\{X_1^j,X_2^j,...,X_{h-1}^j,X_h^j\}$, is utilized as the input of the classification model. This classification model for unobservable customers is developed using the multinomial logistic regression (MLR) algorithm. Compared to other binary classification methods such as random forests, MLR is able to obtain the likelihood of different typical profiles for customers rather than picking a single consumption pattern from the databank \cite{DV2016}. The probability that the $j$'th customer follows the $z$'th typical load profile can be written as \cite{ZX2016}:

\begin{equation}
\label{eq:MLR3}
P(C_j=z|X^j) = \frac{exp(w_z^TX^j)}{\sum_{j=1}^kexp(w_j^TX^j)}
\end{equation}
where, $C_j$ represents the class of the $j$'th unobservable customer, $T$ is the transposition operator, and $w_z$ is the weight vector corresponding to pattern $z$. The learning parameters $w_z$ are obtained by solving $\nabla_{w_z}J=0$ over the training set, where $J$ is the classification risk function, defined as follows \cite{MLR2005}:
\begin{equation}
\label{eq:MLR4}
J = \sum_{j=1}^M[\sum_{z=1}^kc_j^z(w_z)^TX^j-\log\sum_{z=1}^k\exp((w_z)^TX^j)]
\end{equation}
where, $c_j^z$ is the $j$'th element of $c^z$, which is a binary string representing customer class membership. To maximize the log-likelihood function, $J$, with respect to $w_z$, we need to compute the gradient and Hessian of equation \eqref{eq:MLR4}. Based on the block-structured property of learning parameters and Kronecker product of matrices, the gradient and Hessian of the objective function can be obtained and passed to any gradient-based optimizer to find the maximum a posterior (MAP) estimation of model parameters \cite{KM2012}. In this paper, an iterative reweighted least squares (IRLS) training mechanism was implemented \cite{TM2003}. It should be noted that although there are other methods for performing this maximization, none clearly outperforms IRLS \cite{MLR2005}.

\begin{figure}[tbp]
\centering
\includegraphics[width=3.5in]{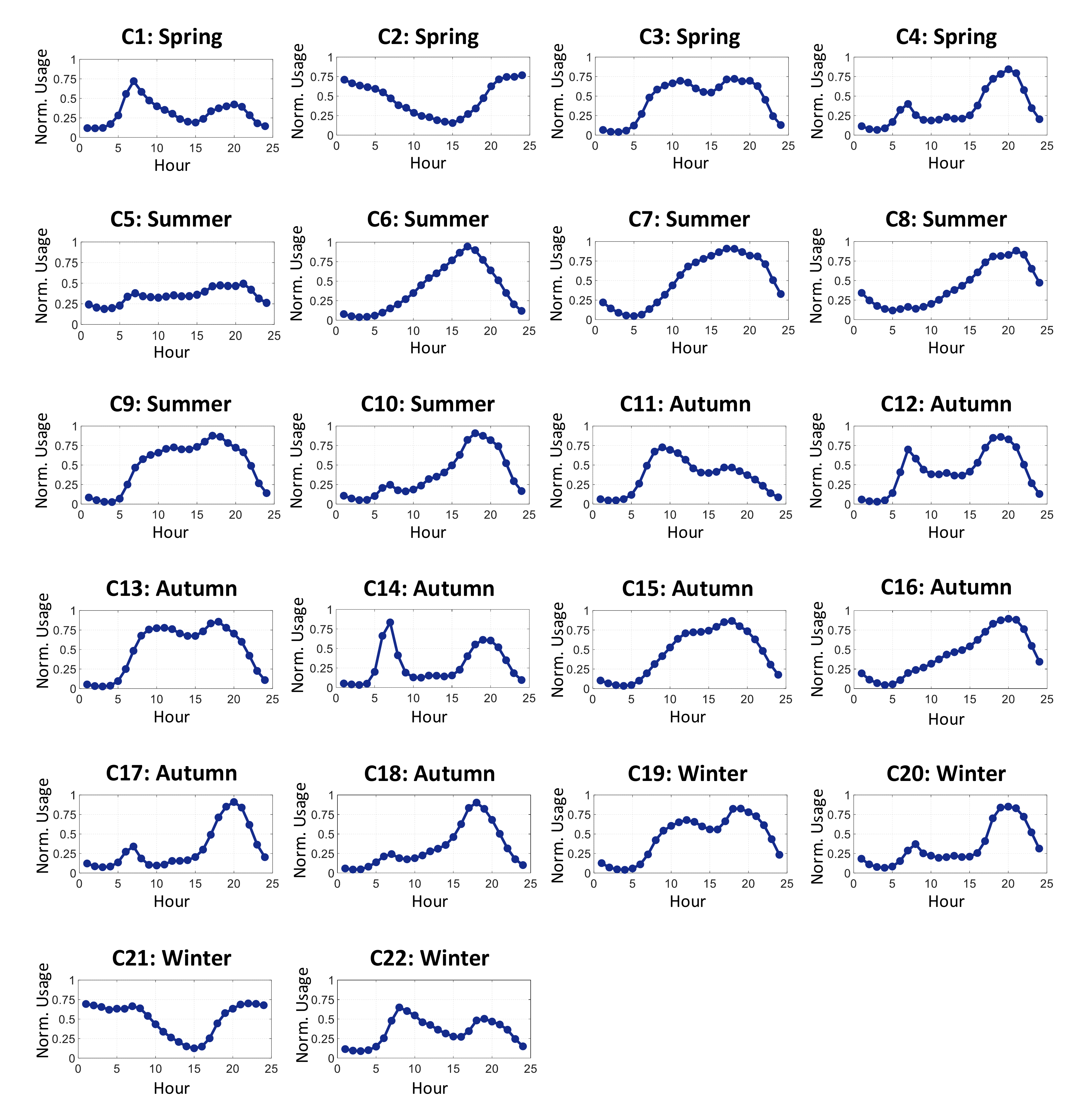}
\caption{Seasonal Typical load patterns databank.}
\label{fig:cluster_result}
\end{figure}
\begin{figure}[tbp]
\centering
\includegraphics[width=3in]{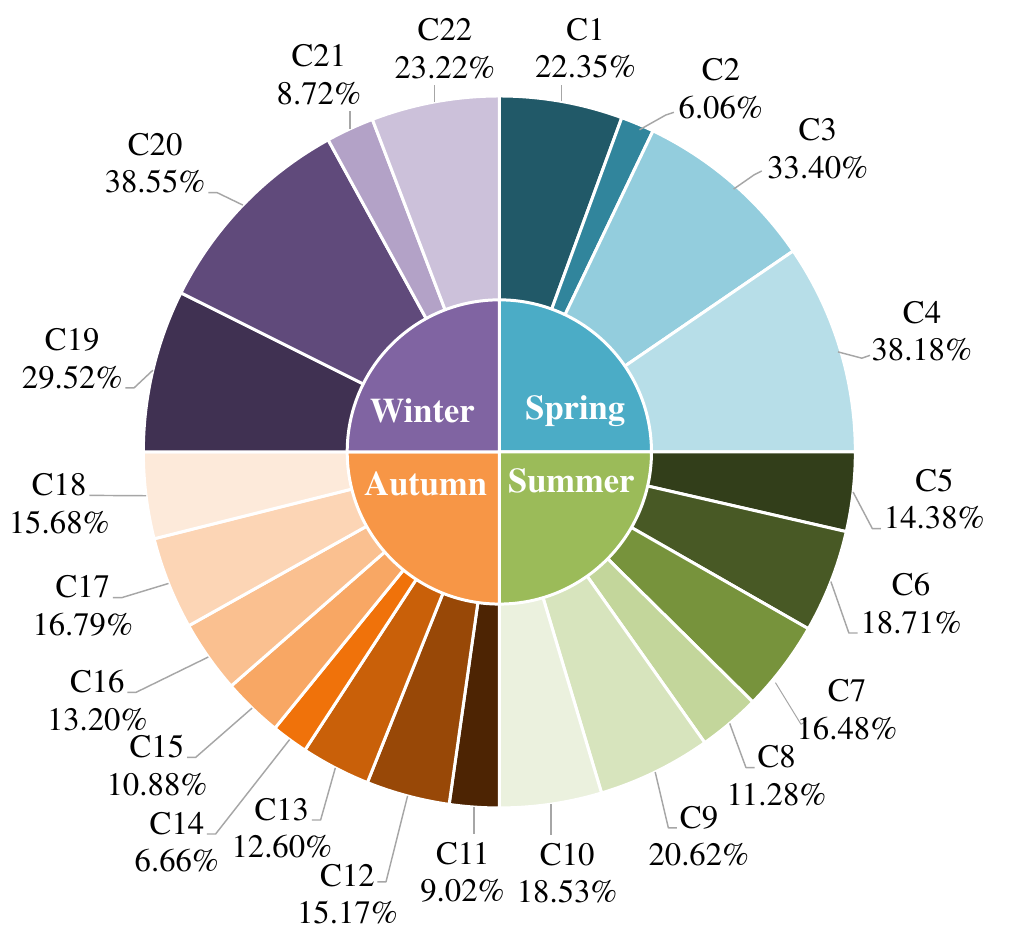}
\caption{Proportion of typical load patterns for different seasons.}
\label{fig:cluster_percentage}
\end{figure}

\subsection{Estimation of CMPC for Unobservable Customers}\label{regression}
To infer the CMPC for unobservable customers, a WCR model is developed by combining two variables: daily load profile and demand level. The basic idea of WCR approach is to utilize the
linear nature of the relationship between the CMPC and monthly energy consumption when the load profiles of customers are similar. This is demonstrated in Fig. \ref{fig:linear}, where the CMPC and monthly energy consumption of customers in different clusters are shown. As depicted in Fig. \ref{fig:linear}, the correlation between monthly energy consumption and the CMPC is largely different for customers with two distinct behavioral patterns in the same season. 

Hence, for $z$'th typical pattern, a linear regression model is trained for mapping the customer's monthly billing information to the CMPC values. The monthly billing data of consumers is obtained by aggregating their SM data. As shown in Fig. \ref{fig:energy_dis} the majority of monthly consumption values are concentrated around 1000 kWh. Then, the actual CMPC value is calculated using the SCADA and SM data at the system peak time. To estimate the parameters $W_z$ and $b_z$ of this regression model, ordinary least square (OLS) is used in this paper \cite{Goodfellow2016}. The basic idea is to minimize the sum of the squares of the differences between the estimated and actual CMPCs. The objective function can be written as follows:
\begin{equation}
\label{eq:objective}
f_z = min_{W_z,b_z}\sum_{i=1}^n(F_{j,m}^i-(E_{j,m}^iW_z+b_z))^2
\end{equation}
where, $E_{j,m}$ and ${F}_{j,m}$ are the monthly consumption level and the actual CMPC for the $j$'th customer at the $m$'th month. It should be noted that our dataset includes the real SM measurements of over 3000 residential customer and the corresponding SCADA records over 3 years. For each regression model, to reduce the overfitting risk, the dataset is randomly divided into two separate subsets for training (80$\%$ of the total data) and testing (20$\%$ of the total data). After training, all regression models are then merged into a WCR to estimate the CMPC for unobservable residential customers. Using the cluster probability values obtained from the classification model, $P(C_j=z|X^j)$, the estimated CMPC for the $j$'th customer at the $m$'th month, $\hat{F}_{j,m}$, is determined as follows:

\begin{equation}
\label{eq:regression}
\hat{F}_{j,m}=\sum_{z=1}^kP(C_j=z|X^j)(W_zE_{j,m}+b_z)
\end{equation}
Hence, the proposed WCR is able to estimate the CMPC of unobservable customers using only their measured monthly consumption within a probabilistic classification setting. OLS regression can produce unbiased estimates that have the smallest variance among all possible linear estimators if the model follows several basic assumptions to satisfy the conditions of Gauss-Markov theorem \cite{GJ2013}. In our work, the linear nature of the relationship between the CMPC and monthly energy consumption in the same cluster and random selection of training data help satisfy these assumptions, thus ensuring the theoretical performance of WCR. Also, it should be noted that in general the performance of the OLS is impacted by outliers and extreme observations \cite{Goodfellow2016}. However, in our problem outliers and extreme values are highly unlikely since the residential customers' monthly demand levels are concentrated within a small range; almost 80$\%$ of customers have monthly consumption levels below 1000 kWh.

\section{Numerical Results}\label{result}
The real distribution system provided by our utility collaborator is equipped with SMs, thus fully observable. This enables us to calculate the exact CMPC of each customer. To test the proposed customer segmentation method for partially observable systems, we assume that 20$\%$ of customers are unobservable and then compare the estimation results with the actual CMPCs. Thus, the data of observable customers (the remaining 80$\%$ of the total data) is divided into 4 subsets corresponding to different seasons of the year for model training.

\subsection{SC Algorithm Performance}
For every subset, the optimal cluster number is determined using DBI and typical load patterns are obtained employing the SC algorithm (detailed in Section \ref{cluster}). Fig. \ref{fig:cluster_result} and Fig. \ref{fig:cluster_percentage} present the $22$ typical load shapes, namely $C_1$, $C_2$, ..., $C_{22}$, and the distribution of population of customers belonging to each cluster during all the seasons. As shown in the figures, the number of typical load profiles in different seasons is not the same and the SC approach is able to capture the critical seasonal consumption patterns. In spring, around $22\%$ of customers show typically higher consumption levels during the morning (around 7:00 am). In contrast, more than $38\%$ of customers have higher energy consumption during the evening (around 20:00 pm). Meanwhile, more than half of customers present low energy consumption value during the afternoon period. The typical load profiles in summer are different from spring. Except for $C_5$, the typical load patterns of $85\%$ of all customers show similar behavioral tendencies. This could be due to air-conditioning load consumption during time intervals with higher temperature. Based on the typical load patterns, the majority of peak demand occurs during the evening interval. For around $74\%$ of customers in summer, the peak time ranges from 17:00 pm to 19:00 pm. In fall, the number of typical load patterns is relatively larger rather than other seasons due to variability of customer behavior. Compared to summer, when peak demand barely happens in the morning, more than $40\%$ of customers have high consumption at around 7:00 am in fall, such as $C_{11}, C_{12}, C_{13}$ and $C_{14}$. Also, around $23\%$ of customers provide almost zero consumption from 10:00 am to 15:00 pm, and nearly one-third of customers show two peaks in the morning and evening periods. The winter typical daily patterns are similar to the results of spring since these two seasons have similar weather in mid-west U.S. 
\begin{figure}[tbp]
\centering
\includegraphics[width=3.5in]{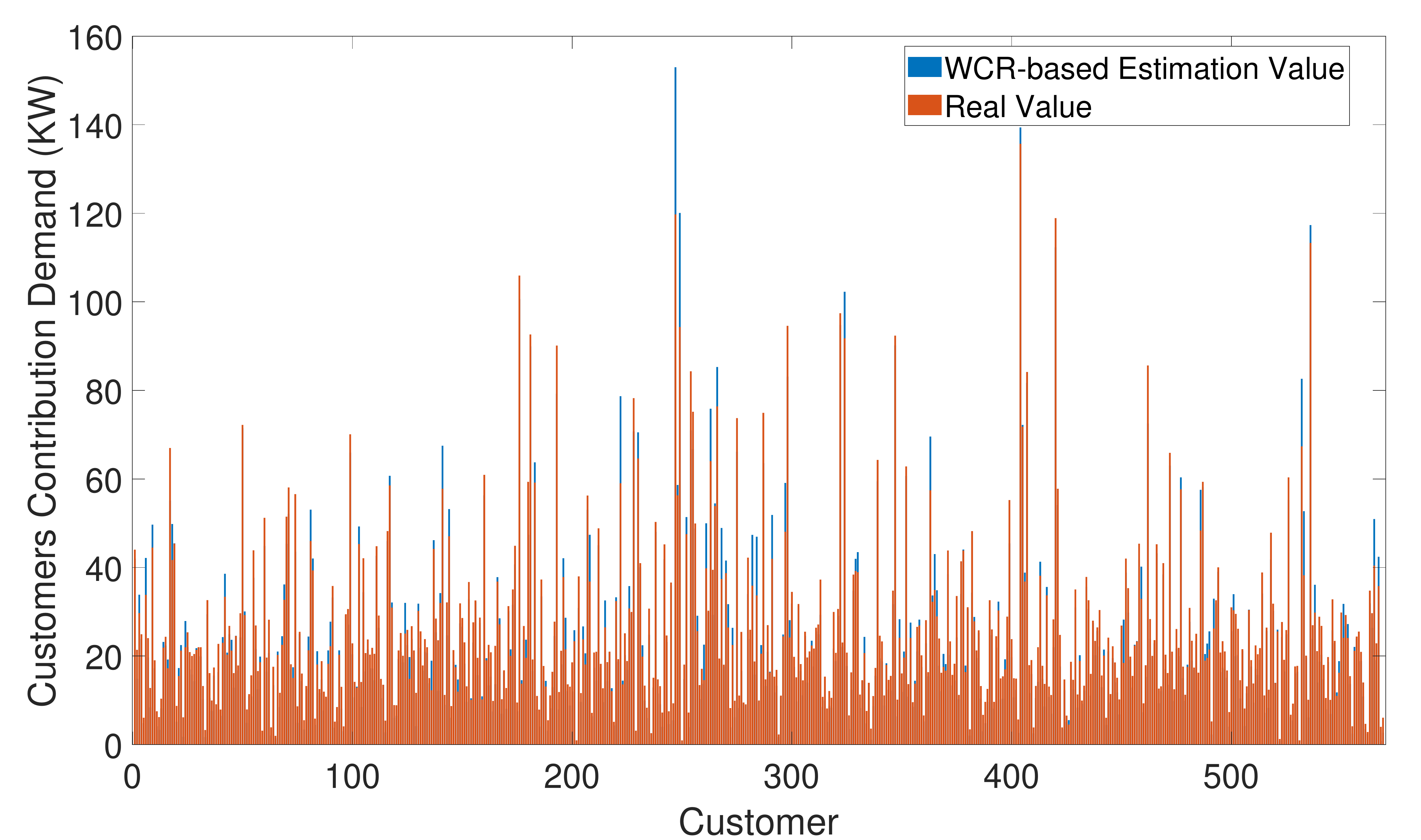}
\caption{Comparison of WCR-based estimation value and real value.}
\label{fig:regression_result}
\end{figure}
\begin{table}[tbp]
\caption{Performance of seasonal WCR models with $R^2$ and MAPE.}
\centering
\setlength{\tabcolsep}{5mm}{
\begin{tabular}{ccc}
\hline\hline
Season & Average $R^2$ & Average MAPE\\[2pt]
\hline
Spring & 0.9446 & 12.44$\%$\\[1pt]
Summer & 0.9071 & 14.24$\%$\\[1pt]
Fall & 0.9384 & 13.18$\%$\\[1pt]
Winter & 0.9204 & 13.7$\%$\\[1pt]
\hline
\end{tabular}}
\label{table:1.1}
\end{table}
%\begin{figure}[htbp]
%\centering
%\subfloat [The histogram of $R^2$ value]{
%\includegraphics[width=3in]{figure/R_2_hist.pdf}
%}
%\hfill
%\subfloat [The histogram of MAPE]{
%\includegraphics[width=3in]{figure/mape_hist.pdf}
%}
%\caption{WCR model performance (spring subset).}
%\label{fig:spring_WCR}
%\end{figure}

\subsection{WCR Performance}
When the seasonal consumption pattern bank is developed using the SM data of observable customers, the WCR models are utilized to infer the CMPC of unobservable customers. 

\subsubsection{Classification Performance Analysis}
For the classification part, the Area under the Curve (AUC) index is employed to assess the performance of MLR model \cite{AUG1982}. AUC is determined as follows:
\begin{equation}
\label{eq:AUC}
\gamma=\int_0^1\frac{TP}{TP+FN}d\frac{FP}{FP+TN}=\int_0^1\frac{TP}{P}d\frac{FP}{N}
\end{equation}
where, TP is the True Positive, TN is the True Negative, FP is the False Positive, FN is the False Negative, and N is the number of total Negatives. Compared to the commonly-used metric, accuracy, the AUC does not depend on the cut-off value that is applied to the posterior probabilities to evaluate the performance of a classification model \cite{DT2012}.

The meaningful range of AUC is between 0.5 to 1. In order to avoid the overfitting problem, the $k$-fold cross-validation method is applied to the MLR to ensure the randomness of the training set \cite{TG1998}. Based on the prior information on customer peak timing distribution, the MLR achieves an AUC value of 0.7 when assigning daily load patterns to unobservable customers.

\subsubsection{Regression Performance Analysis}

Based on the WCR approach, the CMPC of unobservable customers can be estimated using the monthly billing data. Fig. \ref{fig:regression_result} shows the performance of WCR by comparing the actual CMPC with the estimated CMPC for each customer in the testing set for one month. As can be seen, the estimated values are able to accurately track the unobservable customer's real contribution to system peak demand. To assess the performance of the model, the goodness-of-fit measure, $R^2$, and the mean absolute percentage error (MAPE) are utilized in this paper. These two indices are presented in Table \ref{table:1.1} for all seasons. Based on these results, the regression model has a good performance for estimation of CMPC of unobservable customers in this case. 
\begin{figure}[tbp]
\centering
\includegraphics[width=3.5in]{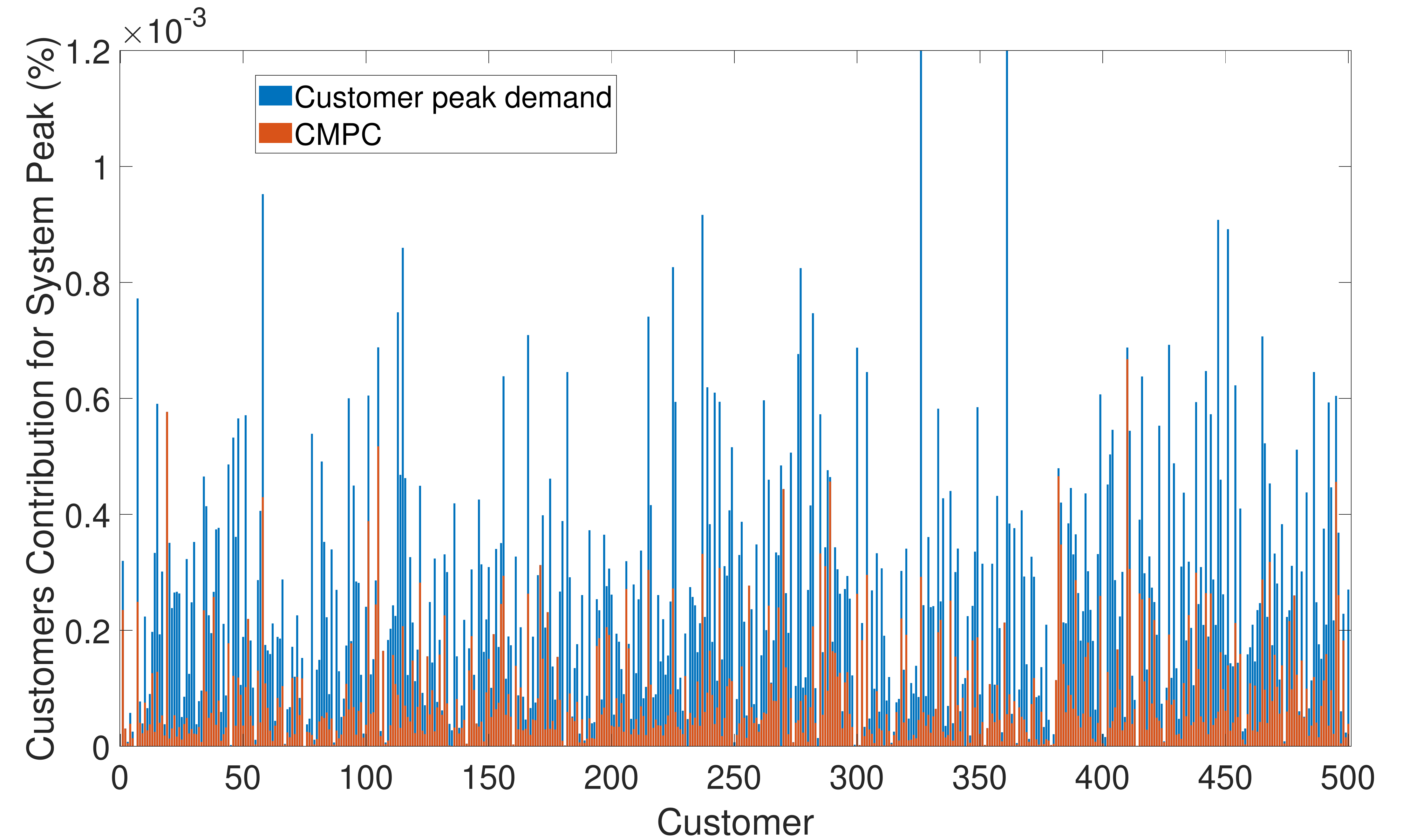}
\caption{Comparison of CMPC and customer peak demand.}
\label{fig:compare_ind_1}
\end{figure}
\begin{figure}[tbp]
\centering
\includegraphics[width=3.5in]{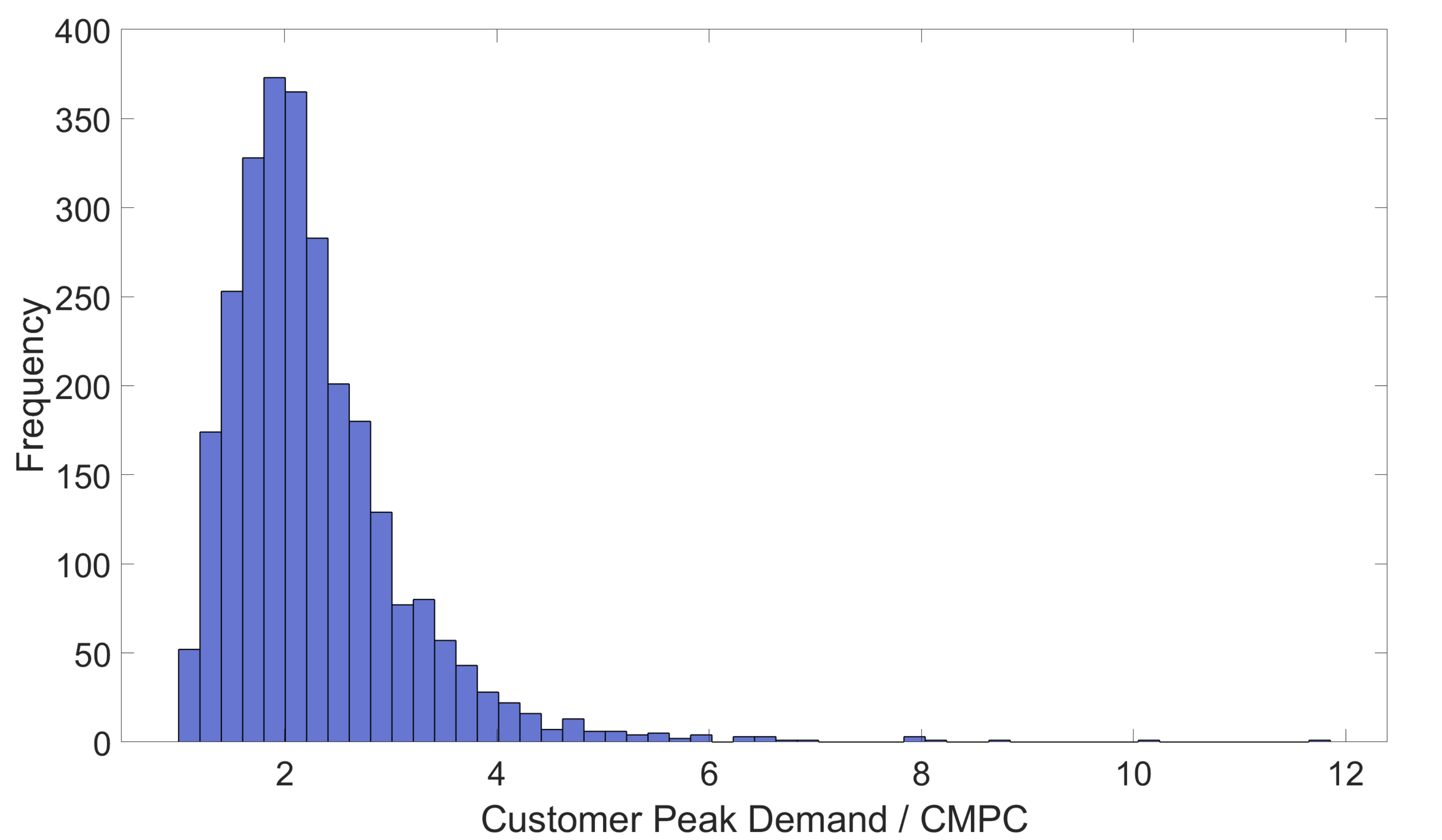}
\caption{The histogram of customer peak demand over CMPC ratio.}
\label{fig:compare_ind_2}
\end{figure}
%\subsection{DR Potential Assessment Metric Comparison}
\subsection{Metric and Method Comparison}
In this section, we demonstrate that the proposed segmentation strategy can target suitable customers, which cannot be classified by existing method in the literature, including customer peak demand-based and load profile entropy-based segmentation strategies \cite{JK2014,RL20152}. Furthermore, to validate the performance of our multi-stage machine learning framework, we have compared the peak contribution estimation MAPE of the proposed learning-based framework with previous method \cite{KW2016}.

\textit{1) Comparing customer peak demand-based strategy and proposed method}: Customer peak demand is a conventional index to describe the potential impact of individual customers on the overall peak demand, which is commonly-used by utilities to perform customer segmentation \cite{JK2014}. In Fig. \ref{fig:compare_ind_1}, the difference between the proposed CMPC and customer peak demand values are presented. It can be seen that the customer peak demand values are generally much higher than CMPC values due to the diversity of load behaviors. According to Fig. \ref{fig:compare_ind_2}, the customer's peak demand can reach five times the customer's actual contribution to the system peak. This considerable difference shows that compared to the proposed method, customer peak demand-based strategy is a very conservative method of quantifying the actual impact of customers, which could lead to unnecessary over-investments in AMI expansion. 

\begin{figure}[tbp]
\centering
\includegraphics[width=3.5in]{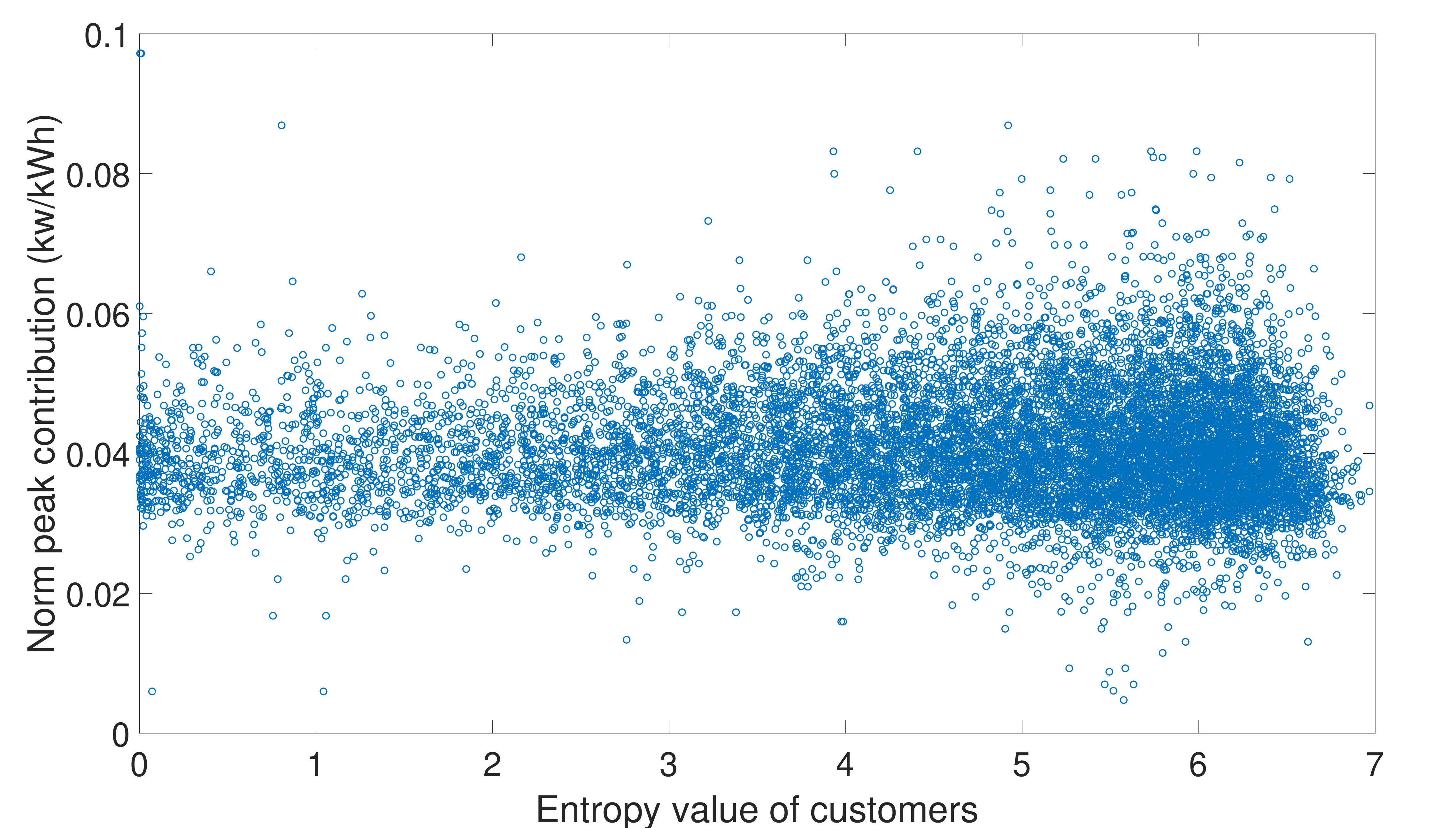}
\caption{The relationship between CMPC and entropy.}
\label{fig:entropy}
\end{figure}

\textit{2) Comparing load profile entropy-based strategy and proposed method}: Entropy is a measure of the variability and uncertainty of customer demand, which has been used to develop customer segmentation approach for peak shaving program targeting \cite{RL20152}.  Customers with lower entropy levels have stable consumption behaviors, which makes them higher priority candidates for peak reduction. In Fig. \ref{fig:entropy}, the relationship between CMPC and entropy is presented. It is observable that customers with high CMPC do not necessarily have low entropy values. This indicates that these two concepts are almost uncorrelated and do not contain mutual information. Hence, unlike the proposed method, the entropy-based strategy does not provide information about customers' impact on system peak demand, and thus, cannot be used as a generic strategy for guiding peak shaving/AMI planning.

\begin{figure}[tbp]
\centering
\includegraphics[width=3.5in]{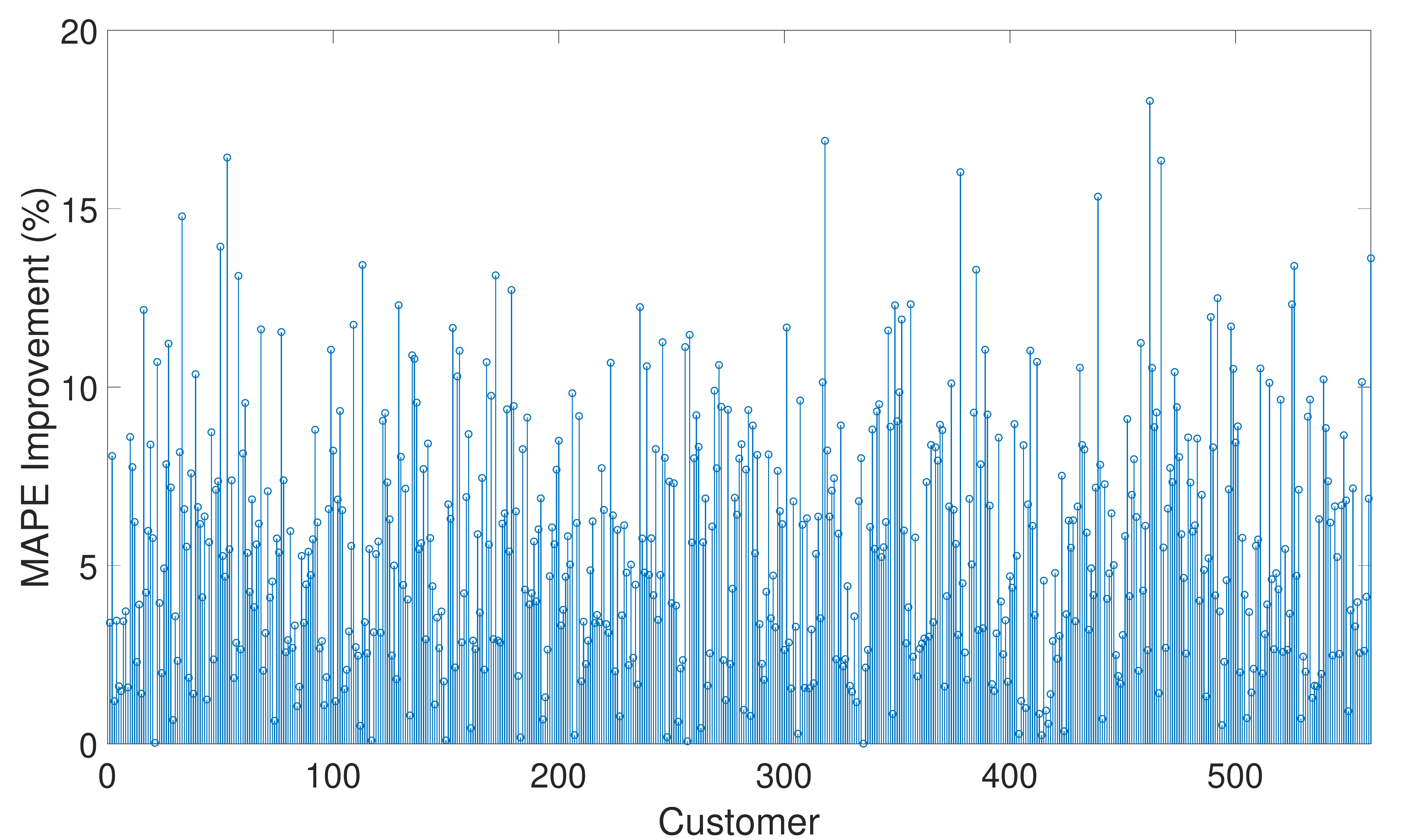}
\caption{Comparison of proposed method and existing method \cite{KW2016}.}
\label{fig:compare}
\end{figure}

\textit{3) Comparing the performance of the proposed multi-stage machine learning-based framework with an existing method}: The performance of the proposed multi-stage machine learning framework is compared with an existing baseline method \cite{KW2016} in terms of estimation accuracy. The baseline method uses ordinary least square regression to determine the peak demand based on the periodic energy consumption. As shown in Fig. \ref{fig:compare}, the estimation MAPE values for our proposed method are generally lower than the results obtained from the previous method in \cite{KW2016}. Our framework has been able to improve the estimation MAPE by $5\%$ on average. Furthermore, a maximum point-wise improvement level of $18\%$ has been achieved over the previous baseline method. Hence, based on this AMI dataset, the proposed method shows a better estimation accuracy compared to the previous work.


\subsection{Application of the Proposed Metric and Strategy}
\begin{figure}[tbp]
\centering
\includegraphics[width=3.5in]{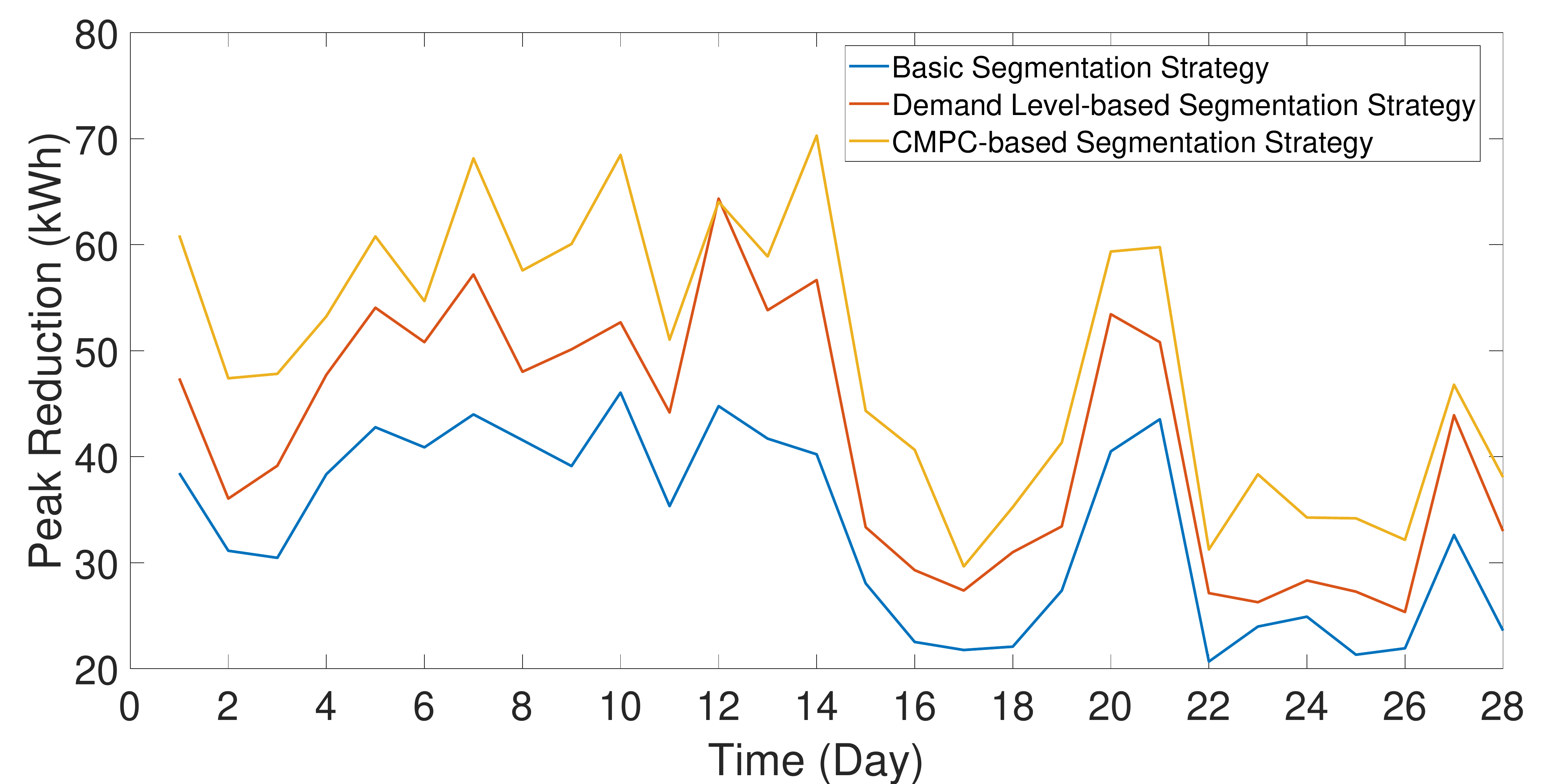}
\caption{Comparison of peak reduction using three different segmentation strategies.}
\label{fig:dr_case}
\end{figure}

To evaluate the performance of the proposed metric and the strategy in system operation, we have applied our works to a basic direct load control-based demand response model, which gives utilities the option to remotely shut down appliances during daily peak demand periods \cite{cc2014}. A 300-house radial distribution network has been considered to evaluate the performance of different segmentation strategies. $35\%$ of unobservable customers are selected for meter installation and participation in peak shaving using three different segmentation metrics: 1) select residential candidates randomly (base strategy); 2) select residential candidates by ranking monthly demand level; 3) select residential candidates based on the proposed CMPC. According to the existing works \cite{dr2013,cv2014}, we have assumed average load elasticity of customers to be 0.21 p.u. We have the compared daily peak reductions in one month (28 days) under the three different customer segmentation strategies. As shown in Fig. \ref{fig:dr_case}, using the proposed CMPC strategy, over 1400$kWh$ peak demand has been saved in this month, which is higher than the other two segmentation strategies. Specifically, in this case, when basic and demand level-based strategies are replaced by CMPC-based strategy, the average peak reduction increases by 50.4$\%$ and 19.7$\%$, respectively. Thus, by comparison, the proposed customer segmentation strategy and the CMPC metric have the potential to provide enhanced customer targeting guidelines for improving operational frameworks. As a future research direction, we will utilize the proposed metric in more advanced and detailed operation models.

\section{Conclusion}\label{conclusion}
In this paper, we have presented a new metric for customer segmentation, CMPC, which can quantify the contributions of individual customers to system peak demand. Moreover, to accurately estimate the CMPC of unobservable residential customers, an innovative three-stage machine learning framework is developed using only their monthly billing data. Employing our real SM data, it is demonstrated and validated that the proposed metric provides utilities with additional actionable information for customer segmentation compared to the existing metrics. This segmentation strategy helps utilities effectively identify impactful customers from thousands of unobservable customers for investment decisions, such as AMI expansion. Also, these customers can be targeted as candidates for residential-level demand-side management (DSM) programs to reduce the critical system peak demand, thus, decreasing the high marginal cost and the risk of system failure. Our work offers other potential benefits for utilities. For example, recently, utilities have been showing increasing interest in residential-level retail price design due to the significant contribution of residential customers to the system peak. The proposed CMPC, together with the developed machine learning framework, can provide a reasonable strategy to obtain guidelines for retail price design by accurately quantifying the impact of residential customers on the system.

\ifCLASSOPTIONcaptionsoff
  \newpage
\fi



\bibliographystyle{IEEEtran}
\bibliography{IEEEabrv,./bibtex/bib/IEEEexample}

\begin{thebibliography}{10}
\providecommand{\url}[1]{#1}
\csname url@samestyle\endcsname
\providecommand{\newblock}{\relax}
\providecommand{\bibinfo}[2]{#2}
\providecommand{\BIBentrySTDinterwordspacing}{\spaceskip=0pt\relax}
\providecommand{\BIBentryALTinterwordstretchfactor}{4}
\providecommand{\BIBentryALTinterwordspacing}{\spaceskip=\fontdimen2\font plus
\BIBentryALTinterwordstretchfactor\fontdimen3\font minus
  \fontdimen4\font\relax}
\providecommand{\BIBforeignlanguage}[2]{{%
\expandafter\ifx\csname l@#1\endcsname\relax
\typeout{** WARNING: IEEEtran.bst: No hyphenation pattern has been}%
\typeout{** loaded for the language `#1'. Using the pattern for}%
\typeout{** the default language instead.}%
\else
\language=\csname l@#1\endcsname
\fi
#2}}
\providecommand{\BIBdecl}{\relax}
\BIBdecl

\bibitem{RG2015}
R.~Gulbinas, A.~Khosrowpour, and J.~Taylor, ``Segmentation and classification
  of commercial building occupants by energy-use efficiency and
  predictability,'' \emph{IEEE Trans. Smart Grid}, vol.~6, no.~3, pp.
  1414--1424, May 2015.

\bibitem{yuan2019}
Y.~{Yuan}, K.~{Dehghanpour}, F.~{Bu}, and Z.~{Wang}, ``A multi-timescale
  data-driven approach to enhance distribution system observability,''
  \emph{IEEE Trans. Power Syst.}, vol.~34, no.~4, pp. 3168--3177, Jul. 2019.

\bibitem{eiasmart}
\BIBentryALTinterwordspacing
{Energy Information Administration}. (2017) Advanced metering count by
  technology type. [Online]. Available:
  \url{https://www.eia.gov/electricity/annual/html/epa_10_10.html}
\BIBentrySTDinterwordspacing

\bibitem{cluster2013}
M.~Koivisto, P.~Heine, I.~Mellin, and M.~Lehtonen, ``Clustering of connection
  points and load modeling in distribution systems,'' \emph{IEEE Trans. Power
  Syst.}, vol.~28, no.~2, pp. 1255--1265, May 2013.

\bibitem{cluster2016}
S.~Haben, C.~Singleton, and P.~Grindrod, ``Analysis and clustering of
  residential customers energy behavioral demand using smart meter data,''
  \emph{IEEE Trans. Smart Grid}, vol.~7, no.~1, pp. 136--144, Jan. 2016.

\bibitem{RL20152}
R.~Li, C.~Gu, F.~Li, G.~Shaddick, and M.~Dale, ``Development of low voltage
  network templates part ii: Peak load estimation by clusterwise regression,''
  \emph{IEEE Trans. Power Syst.}, vol.~30, no.~6, pp. 3045--3052, Nov. 2015.

\bibitem{AK2015}
A.~Kavousian, R.~Rajagopal, and M.~Fischer, ``Ranking appliance energy
  efficiency in households: Utilizing smart meter data and energy efficiency
  frontiers to estimate and identify the determinants of appliance energy
  efficiency in residential buildings,'' \emph{Energy and Buildings}, vol.~99,
  pp. 220--230, Apr. 2015.

\bibitem{JK2014}
J.~Kwac, J.~Flora, and R.~Rajagopal, ``Household energy consumption
  segmentation using hourly data,'' \emph{IEEE Trans. Smart Grid}, vol.~5,
  no.~1, pp. 420--430, Jan. 2014.

\bibitem{good2018}
Y.~Yu, G.~Liu, W.~Zhu, F.~Wang, B.~Shu, K.~Zhang, N.~Astier, and R.~Rajagopal,
  ``Good consumer or bad consumer: Economic information revealed from demand
  profiles,'' \emph{IEEE Trans. Smart Grid}, vol.~9, no.~3, pp. 2347--2358, May
  2018.

\bibitem{MYS2019}
M.~{Sun}, Y.~{Wang}, G.~{Strbac}, and C.~{Kang}, ``Probabilistic peak load
  estimation in smart cities using smart meter data,'' \emph{IEEE Trans. Ind.
  Electron.}, vol.~66, no.~2, pp. 1608--1618, Feb. 2019.

\bibitem{pss}
M.~Uddin, M.~F. Romlie, M.~F. Abdullah, S.~A. Halim, A.~H.~A. Bakar, and T.~C.
  Kwang, ``A review on peak load shaving strategies,'' \emph{Renewable and
  Sustainable Energy Reviews}, vol.~82, pp. 3323--3332, 2018.

\bibitem{wzy}
\BIBentryALTinterwordspacing
Z.~Wang. Dr. {Zhaoyu Wang's} home page. [Online]. Available:
  \url{http://wzy.ece.iastate.edu/Testsystem.html}
\BIBentrySTDinterwordspacing

\bibitem{zscore}
D.~Cousineau and S.~Chartier, ``Outlier detection and treatment: a review,''
  \emph{International Journal of Psychological Research}, vol.~3, no.~1, pp.
  58--67, Jan. 2010.

\bibitem{RL2015}
R.~Li, C.~Gu, F.~Li, G.~Shaddick, and M.~Dale, ``Development of low voltage
  network templates—part ii: Peak load estimation by clusterwise
  regression,'' \emph{{IEEE} Trans. Power Syst.}, vol.~30, no.~6, pp.
  3045--3052, Nov. 2015.

\bibitem{KC2017}
K.~Chen, J.~Hu, and Z.~He, ``Data-driven residential customer aggregation based
  on seasonal behavioral patterns,'' \emph{2017 IEEE Power Energy Society
  General Meeting}, pp. 1--5, Jul. 2017.

\bibitem{NMW2002}
A.~Ng, M.~Jordan, and Y.~Weiss, ``On spectral clustering: analysis and an
  algorithm,'' \emph{Advances in Neural Information Processing Systems}, pp.
  849--856, 2002.

\bibitem{kaveh2019}
K.~Dehghanpour, Y.~Yuan, Z.~Wang, and F.~Bu, ``A game-theoretic data-driven
  approach for pseudo-measurement generation in distribution system state
  estimation,'' \emph{IEEE Trans. Smart Grid}, pp. 1--1, 2019.

\bibitem{GJ2007}
U.Luxburg, ``A tutorial on spectral clustering,'' \emph{Statistics and
  Computing}, vol.~17, no.~4, pp. 395--416, Mar. 2007.

\bibitem{ISD2007}
I.~S. Dhillon, Y.~Guan, and B.~Kulis, ``Weighted graph cuts without
  eigenvectors a multilevel approach,'' \emph{{IEEE} Trans. Pattern Anal.
  Machine Intell.}, vol.~29, no.~11, pp. 1944--1957, Nov. 2007.

\bibitem{lap1997}
F.~R.~K. Chung, \emph{Spectral Graph Theory}.\hskip 1em plus 0.5em minus
  0.4em\relax American Mathematical Society, 1997.

\bibitem{DS1996}
D.~Spielman and S.~Teng, ``Spectral partitioning works: Planar graphs and
  finite element meshes,'' \emph{In Proceedings of the 37th Annual Symposium on
  Foundations of Computer Science}, pp. 1--1, 1996.

\bibitem{CC2013}
C.~C. Aggarwal and C.~K. Reddy, \emph{Data Clustering: Algorithms and
  Applications}.\hskip 1em plus 0.5em minus 0.4em\relax Chapman $\&$ Hall/CRC,
  2013.

\bibitem{DV2016}
D.~Vercamer, B.~Steurtewagen, D.~V. den Poel, and F.~Vermeulen, ``Predicting
  consumer load profiles using commercial and open data,'' \emph{IEEE Trans.
  Power Syst.}, vol.~31, no.~5, pp. 3693--3701, Sep. 2016.

\bibitem{ZX2016}
Z.~Xu, Z.~Hong, Y.~Zhang, J.~Wu, A.~C. Tsoi, and D.~Tao, ``Multinomial latent
  logistic regression for image understanding,'' \emph{{IEEE} Trans. on Image
  Process.}, vol.~25, no.~2, pp. 973--987, Feb. 2016.

\bibitem{MLR2005}
B.~Krishnapuram, L.~Carin, M.~A.~T. Figueiredo, and A.~J. Hartemink, ``Sparse
  multinomial logistic regression: Fast algorithms and generalization bounds,''
  \emph{IEEE Trans. Pattern Anal. Mach. Intell.}, vol.~27, no.~6, pp. 957--968,
  Jun. 2005.

\bibitem{KM2012}
K.~P. Murphy, \emph{Machine Learning: A Probabilistic Perspective}.\hskip 1em
  plus 0.5em minus 0.4em\relax MIT Press, 2012.

\bibitem{TM2003}
T.~Minka, ``A comparison of numerical optimizers for logistic regression,''
  \emph{technical report, Dept. of Statistics, Carnegie Mellon Univ.}, pp.
  1--1, 2003.

\bibitem{Goodfellow2016}
I.~Goodfellow, Y.~Bengio, and A.~Courville, \emph{Deep Learning}.\hskip 1em
  plus 0.5em minus 0.4em\relax MIT Press, 2016,
  \url{http://www.deeplearningbook.org}.

\bibitem{GJ2013}
G.~James, D.~Witten, T.~Hastie, and R.~Tibshirani, \emph{An Introduction to
  Statistical Learning: with Applications in R}.\hskip 1em plus 0.5em minus
  0.4em\relax New York: Springer, 2013.

\bibitem{AUG1982}
J.~A. Hanley and B.~J. McNeil, ``The meaning and use of the area under a
  receiver operating characteristic (roc) curve,'' \emph{Radiology}, vol. 143,
  no.~1, pp. 29--36, Apr. 1982.

\bibitem{DT2012}
D.~Thorleuchter and D.~V. den Poel, ``Predicting e-commerce company success by
  mining the text of its publicly-accessible website,'' \emph{Expert Syst.
  Applicat.}, vol.~39, no.~17, pp. 13\,026--13\,034, Dec. 2012.

\bibitem{TG1998}
T.~G. Dietterich, ``Approximate statistical tests for comparing supervised
  classification learning algorithms,'' \emph{Neural Comput.}, vol.~10, no.~7,
  pp. 1895--1923, Oct. 1998.

\bibitem{KW2016}
W.~H. Kersting, \emph{Distribution System Modeling and Analysis}.\hskip 1em
  plus 0.5em minus 0.4em\relax CRC Press, 2016.

\bibitem{cc2014}
C.~Chen, J.~Wang, and S.~Kishore, ``A distributed direct load control approach
  for large-scale residential demand response,'' \emph{{IEEE} Trans. Power
  Syst.}, vol.~29, no.~5, pp. 2219--2228, Sep. 2014.

\bibitem{dr2013}
S.~Gyamfi, S.~Krumdieck, and T.~Urmee, ``Residential peak electricity demand
  response—highlights of some behavioural issues,'' \emph{Renewable and
  Sustainable Energy Reviews}, vol.~25, pp. 71--77, 2013.

\bibitem{cv2014}
C.~Vivekananthan, Y.~Mishra, G.~Ledwich, and F.~Li, ``Demand response for
  residential appliances via customer reward scheme,'' \emph{{IEEE} Trans.
  Smart Grid}, vol.~5, no.~2, pp. 809--820, Mar. 2014.

\end{thebibliography}
\begin{IEEEbiography}[{\includegraphics[width=1in,height=1.25in,clip,keepaspectratio]{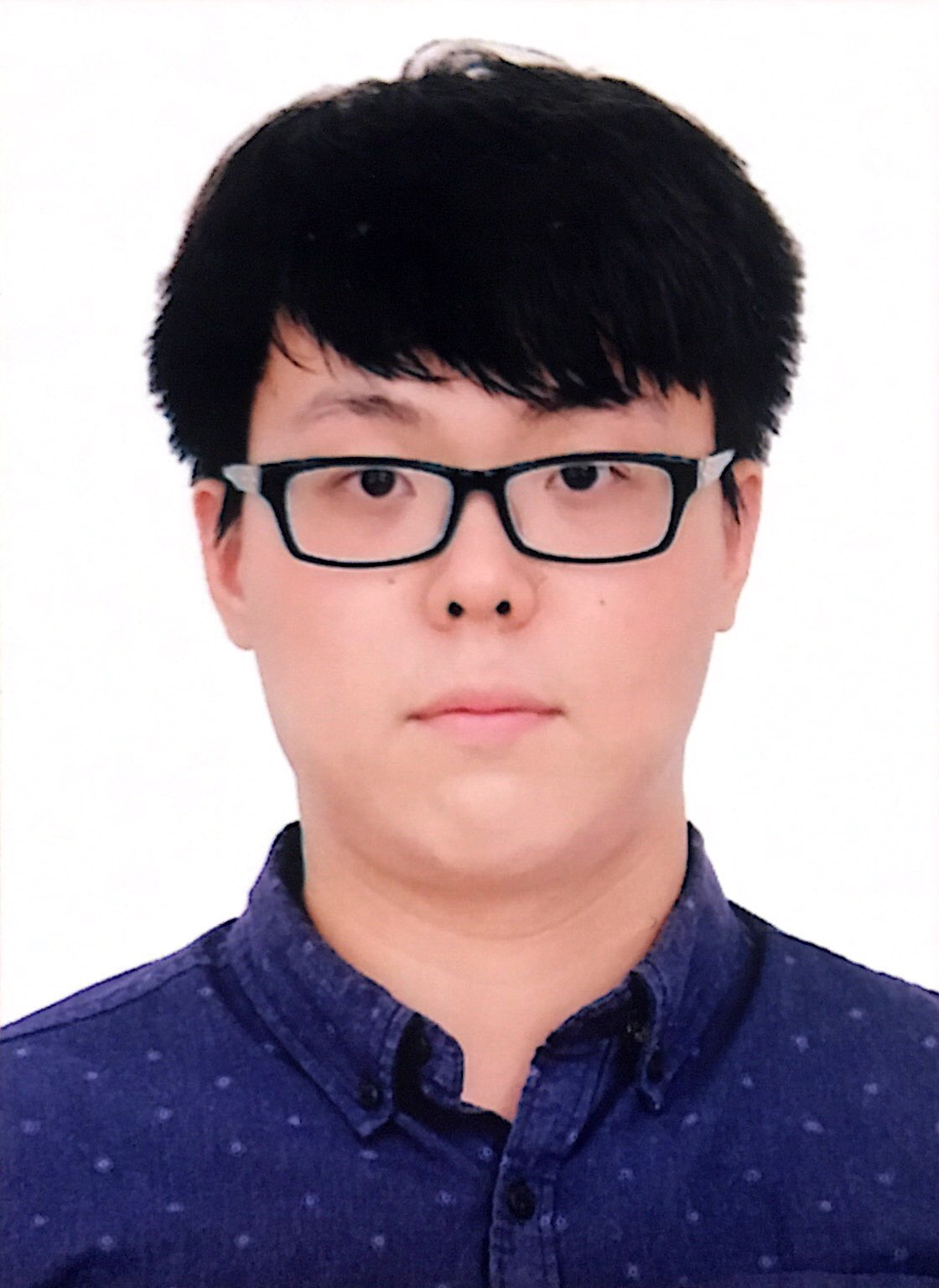}}]{Yuxuan Yuan}(S'18) received the B.S. degree in Electrical \& Computer Engineering from Iowa State University, Ames, IA, in 2017. He is currently pursuing the Ph.D. degree at Iowa State University. His research interests include distribution system state estimation, synthetic networks, data analytics, and machine learning.
\end{IEEEbiography}

\begin{IEEEbiography}[{\includegraphics[width=1in,height=1.25in,clip,keepaspectratio]{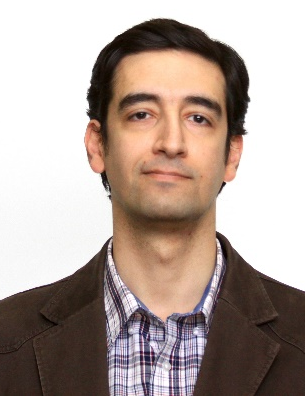}}]{Kaveh Dehghanpour}(S'14--M'17) received his B.Sc. and M.S. from University of Tehran in electrical and computer engineering, in 2011 and 2013, respectively. He received his Ph.D. in electrical engineering from Montana State University in 2017. He is currently a Postdoctoral Research Associate at Iowa State University. His research interests include machine learning and data mining for monitoring and control of smart grids, and market-driven management of distributed energy resources.
\end{IEEEbiography}

\begin{IEEEbiography}[{\includegraphics[width=1in,height=1.25in,clip,keepaspectratio]{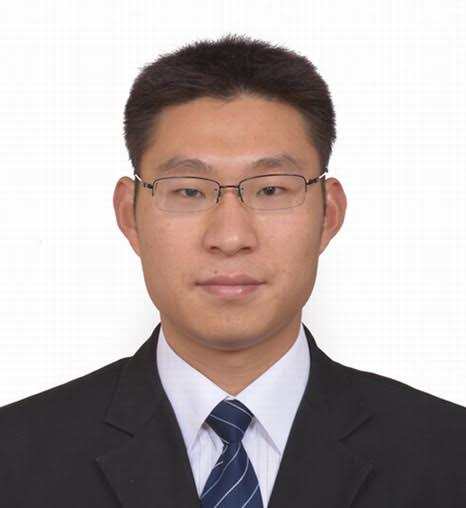}}]{Fankun Bu}(S'18) received the B.S. and M.S. degrees from North China Electric Power University, Baoding, China, in 2008 and 2013, respectively. From 2008 to 2010, he worked as a commissioning engineer for NARI Technology Co., Ltd., Nanjing, China. From 2013 to 2017, he worked as an electrical engineer for State Grid Corporation of China at Jiangsu, Nanjing, China. He is currently pursuing his Ph.D. in the Department of Electrical and Computer Engineering, Iowa State University, Ames, IA. His research interests include distribution system modeling, smart meter data analytics, renewable energy integration, and power system relaying.
\end{IEEEbiography}

\begin{IEEEbiography}[{\includegraphics[width=1in,height=1.25in,clip,keepaspectratio]{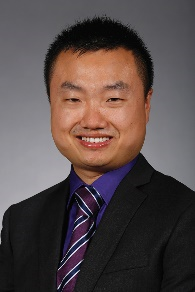}}]{Zhaoyu Wang}(S'13--M'15) is the Harpole-Pentair Assistant Professor with Iowa State University. He received the B.S. and M.S. degrees in electrical engineering from Shanghai Jiaotong University in 2009 and 2012, respectively, and the M.S. and Ph.D. degrees in electrical and computer engineering from Georgia Institute of Technology in 2012 and 2015, respectivelyHis research interests include power distribution systems and microgrids, particularly on their data analytics and optimization. He is the Principal Investigator for a multitude of projects focused on these topics and funded by the National Science Foundation, the Department of Energy, National Laboratories, PSERC, and Iowa Energy Center. Dr. Wang is the Secretary of IEEE Power and Energy Society (PES) Award Subcommittee, Co-Vice Chair of PES Distribution System Operation and Planning Subcommittee, and Vice Chair of PES Task Force on Advances in Natural Disaster Mitigation Methods. He is an editor of IEEE Transactions on Power Systems, IEEE Transactions on Smart Grid, IEEE PES Letters and IEEE Open Access Journal of Power and Energy, and an associate editor of IET Smart Grid.
\end{IEEEbiography}
\end{document}